\documentclass[reqno]{amsart}
\usepackage[%
  paperwidth=192mm,
  paperheight=262mm,
  vmargin={14mm,15mm},
  hmargin={20mm,20mm},
  headsep=12pt,
  footskip=12pt,
]{geometry}

\usepackage[utf8]{inputenc}
\usepackage{mathtools}
\mathtoolsset{showonlyrefs}

\usepackage{subfigure}
\usepackage[usenames]{color}

\usepackage{algorithm}
\usepackage{algpseudocode}

\usepackage{xcolor}
\usepackage{tikz}
\usepackage{amsaddr} 
\usepackage{nicefrac}
\usepackage{tikz-3dplot}
\usepackage{comment}
\usepackage{amsthm,amsmath,amssymb}
\usepackage{enumitem,caption,graphicx}
\usepackage[colorlinks=true, pdfstartview=FitV, linkcolor=blue,citecolor=blue, urlcolor=blue]{hyperref}

\usepackage{array, babel, booktabs, cite, float, footmisc, kvoptions, multicol, nag, pdf14, pdftexcmds, ragged2e, upref, url, xcolor, xpatch, zref-base}

\usepackage{newtxmath, newtxtext}
\usepackage{amsmath,amssymb}

\newcommand{\uu}{\boldsymbol{\mathfrak{u}}}             
\newcommand{\m}{\mathfrak{m}}
\newcommand{\nn}{\mathfrak{n}}

\newcommand{\s}{\mathbb{S}}

\newcommand{\cs}{\texttt{C}}

\newcommand{\Lsf}{\mathsf{L}}
\newcommand{\Msf}{\mathsf{G}}
\newcommand{\F}{\mathbb{F}}
\newcommand{\A}{\boldsymbol{\mathfrak{U}}}              
\newcommand{\B}{{\mathfrak{M}}}
\newcommand{\x}{\mathbf{X}}              

\newcommand{\D}{\mathcal{D}}
\newcommand{\ee}{\mathfrak{e}}

\newcommand{\bu}{\mathbf{U}}             

\newcommand{\e}{{\varepsilon}}

\newcommand{\fp}{\mathfrak{p}}          
\newcommand{\tr}{\operatorname{tr}}     
\newcommand{\arcosh}{\operatorname{arcosh}} 

\newtheorem{theorem}{Theorem}
\newtheorem{lemma}[theorem]{Lemma}
\newtheorem{corollary}[theorem]{Corollary}
\newtheorem{proposition}[theorem]{Proposition}

\newtheorem{remark}[theorem]{Remark}

\title{On the instability of some upward propagating, exact, nonlinear mountain waves}

\author[CP]{Christian Puntini}
\address{Faculty of Mathematics, University of Vienna, Oskar-Morgenstern-Platz 1, 1090 Vienna,	Austria.}
\email{\href{christian.puntini@univie.ac.at}{christian.puntini@univie.ac.at}}

\begin{document}


\begin{abstract}
Using the short-wavelength instability method, we investigate the linear instability of an exact solution describing upward-propagating mountain waves, derived in A.~Constantin, \emph{J. Phys. A: Math. Theor.} (2023), for a dry and adiabatic compressible flow. Within this approach, the stability problem reduces to the analysis of a system of ordinary differential equations with periodic coefficients along the fluid trajectories, which we study by means of Floquet theory. We show that the Floquet multipliers are of the form $\{1,\mu,\mu^{-1}\}$, the multiplier $1$ being associated with the conservation of potential vorticity, so that stability is decided by a single discriminant, which coincides with that of a Hill equation. For neutral stratification the flow is unstable whenever the wave steepness exceeds the critical value $\frac{1}{3}$, while a weakly stable stratification lowers this threshold to $\frac{1}{3}-\frac{4}{9}\nu^2+O(\nu^4)$, $\nu$ being the ratio between the Brunt--V\"ais\"al\"a frequency and the orbital frequency of the air parcels. The instability is a subharmonic parametric resonance. Since the solution is given in Lagrangian coordinates, these results identify an unstable region in the upper part of the laminar layer, beneath the tropopause, where the wave is steepest.
\end{abstract}

\maketitle




\section{Introduction}
\begin{figure}
    \centering
    \includegraphics[width=\linewidth]{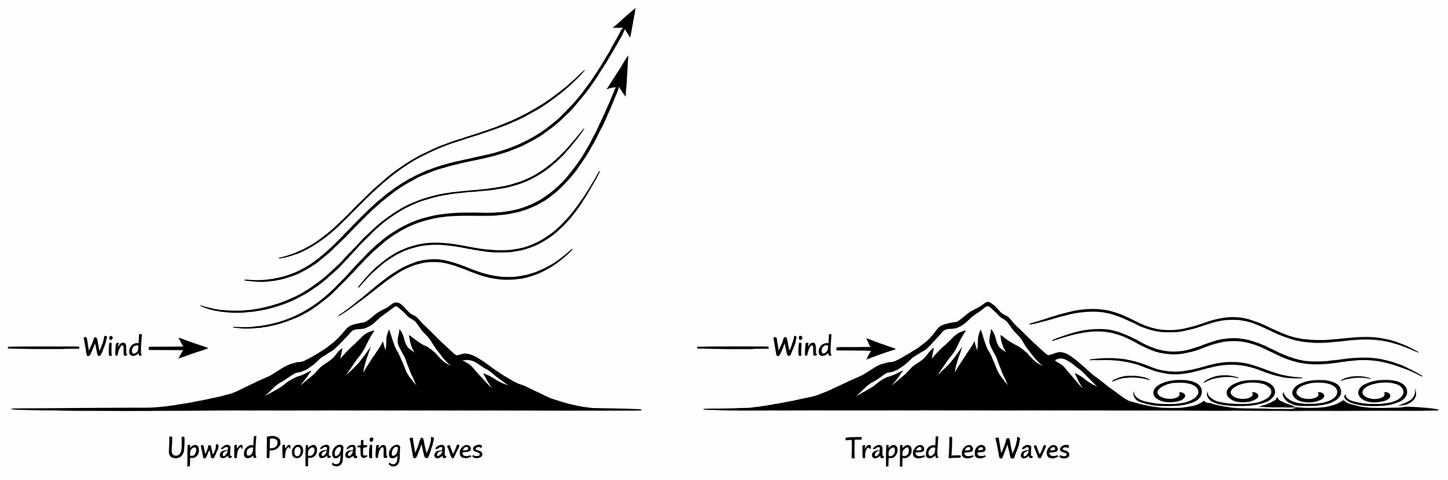}
    \caption{Sketch of the two main types of mountain waves: upward propagating and trapped lee waves. }
    \label{depiction}
\end{figure}

Mountain waves are a specific type of atmospheric gravity wave that form downwind of a long topographic barrier (e.g., a mountain) when a stable, stratified flow with a strong cross-ridge component is forced to ascend over it. As the air rises, buoyancy perturbations develop and generate disturbances that propagate away from the mountain as gravity (buoyancy) waves. Because the atmosphere is compressible and stratified, these waves can propagate both horizontally and vertically (see \cite{durran,durran2}).  As air density decreases with altitude, their amplitude increases, and wave steepening and overturning may occur as the waves penetrate the free atmosphere.\\
While the orographic lift over the ridge is relatively uniform, the flow downstream of the crest is much more complex. It typically features a laminar layer where smooth mountain waves travel, and layers of clear-air turbulence—irregular, high-frequency motions that are not visible to the naked eye. In the lower part of the downstream flow, rotors are often present: rapidly rotating vortices generated by frictional and viscous effects in the descending air near the surface.\\
The meteorological literature distinguishes two main types of mountain waves: vertically propagating waves and trapped lee waves. See Fig.\ref{depiction}. Vertically propagating waves arise when, above the peak, temperature and density decrease with height and wind speed does not increase markedly; their amplitude grows upward and they can produce strong vertical motions (exceeding $30\,\mathrm{m\,s^{-1}}$). If they break before reaching the tropopause, they can generate significant clear-air turbulence, creating important hazards to aviation (see e.g. \cite{Lilly, GuarinoETAL, Bramberg}).\\
Trapped lee waves, by contrast, occur when vertical shear or a low-level inversion creates a waveguide that confines energy near ridge height, yielding a stationary downstream wave train with limited vertical propagation, often marked by stacked lenticular clouds and sometimes accompanied by rotors beneath the first crest (see \cite{teix} and \cite{ConstantinMountain2023,ConstantinWeber2025MountainWaves}). 
\\
\\
In \cite{ConstantinMountain2023}, an explicit and exact solution for mountain waves propagating upwards---in the Lagrangian framework---consisting of oscillations superimposed on a mean current propagating upward has been proposed.  
Although explicit solutions—however complicated they may be—are idealised and therefore approximate representations of reality, they can be a fruitful tool for a better understanding of the phenomenon under analysis, possibly as a starting point for a perturbation analysis (see e.g. \cite{C_JPO, Abrashkin, P_DIE} for exact solutions in oceanography). Once an explicit solution has been obtained, a natural question is whether it is stable or unstable and, in the latter case, under what conditions. This is the scope we pursue in this work.\\
We adopt the short-wavelength instability approach, in the most general description given by \cite{LH} (see also \cite{Bayly, FriedlanderVishik1991}), which has proved to be particularly suited for the instability analysis of explicit solutions in Lagrangian coordinates for oceanic flows (see e.g. \cite{Leblanc2004,CG,Kruse,Kruse2,mioInstability}). It has to be noted however, that, even if the general approach of \cite{LH,LebovitzL} or \cite{FriedlanderVishik1991, Friedlander} has been applied to fluid flows in the Lagrangian setting only in the last decade or so, the analysis of short-wavelength instabilities has been an important, classical subject in the theory of hydrodynamic stability (see e.g. \cite{ES78,ES80,LS83}).\\
This paper is structured as follows: in Section \ref{model} we review the solution proposed in \cite{ConstantinMountain2023}, and then, in Section \ref{Sec:instability}, which is the main subject of this work, we use the short-wavelength instability approach to prove that such a solution is unstable. As the analysis will be based on the study of a system of ordinary differential equations with time-periodic coefficient, we will invoke Floquet theory to study the behaviour of its solutions. In particular, after identifying two suitable parameters (the wave steepness and a stratification parameter) that will characterise the nature of the evolution of the aforementioned solutions, we will recall the basics of Floquet theory. Then we will apply the theory to the problem under consideration by analysing the monodromy matrix of the system of ODEs, and the analysis can be reduced to the study of a Hill equation. For a neutrally stratified fluid we will provide an exact criterion for the onset on the instability, analogous of that of \cite{Leblanc2004} for Gerstner waves. For general stratification, we will provide an instability diagram obtained numerically, which shows regions, in the parameter space, where the solution of \cite{ConstantinMountain2023} is unstable.  We conclude in Section \ref{discussion} with a discussion of the results.

\section{Review of the exact solution in \texorpdfstring{\cite{ConstantinMountain2023}}{Constantin et al. (2023)}}\label{model}
\begin{figure}
            \centering
            \includegraphics[width=.6\linewidth]{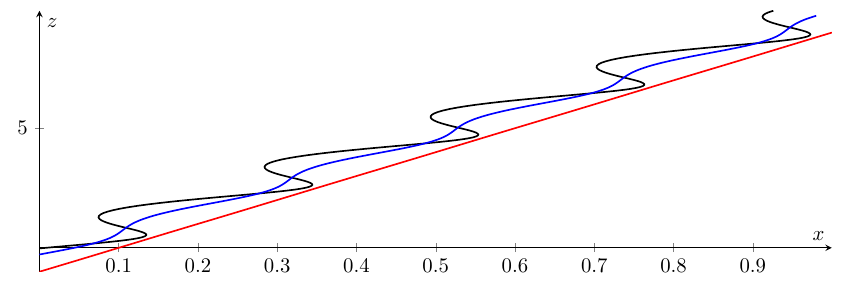}
            \caption{Particle path of \eqref{solution} with $a=s=0$ and $b=-0.1$ (black), $b=-0.3$ (blue), $b=-1$ (red) as time evolves, with $t\in [0,10] $. Moreover, $\bar{u} =0.1$, $\bar{w} =1$ and $k=20/3$. At time $t=0$, the particle initial position is in the bottom-left corner, and at final time $t=10$ its position is in the top right corner. Due to the exponential decrease of the wave amplitude with $b$, for $b=-1$ the oscillations are basically not visible.}
            \label{fig:path}
        \end{figure}

\begin{figure}
            \centering
            \includegraphics[width=.7\linewidth]{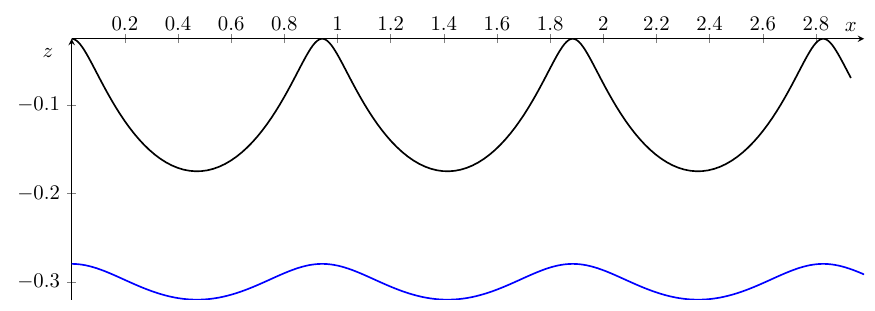}
            \caption{Material lines $b=$const (i.e. isentropes) path of \eqref{solution} at fixed time $t=0$,  with $a \in [0,3]$, $s=0$ and $b=-0.1$ (black), $b=-0.3$ (blue), with  $k=20/3$. For $b=-1$ the oscillations are not visible, so it is not plotted.}
            \label{fig:path2}
        \end{figure}

\noindent The equations of motion for the description of mountain waves are: the Euler equations 
\begin{equation}\label{eq:Euler2}
\begin{aligned}
\frac{Du}{Dt}= -\frac{1}{\rho} P_x, \qquad
\frac{D v}{Dt}= -\frac{1}{\rho} P_y, \qquad\frac{D w}{Dt} &= -\frac{1}{\rho} P_z - g,
\end{aligned}\end{equation}
for a velocity field $\bu =(u,v,w)$, the pressure $P$ and density $\rho$, and where ${D}/{Dt}={\partial}/{\partial t} + \bu \cdot \nabla$ is the material derivative. These are coupled with the equations of: 
\begin{align}\text{mass conservation}&\qquad \rho_t + u\,\rho_x + v\,\rho_y + w\,\rho_z
+ \rho\left( u_x + v_y + w_z \right) = 0;\label{mass} \\
				\text{ state for an ideal gas}&\qquad
                 \label{gas} P=\rho { T};\\
                 \text{first law of thermodynamics}&\qquad
T_t + u\,T_x + v\,T_y + w\,T_z
- \frac{\kappa}{\rho}\left( P_t + u\,P_x + v\,P_y + w\,P_z \right) = 0.\label{eq:thermolaw} \end{align}
Here, $T$ is the temperature, and the equations have already been put into non-dimensional form, to which dimensional quantities (with primes) are related by
\begin{equation}\begin{aligned}
	t' &= \frac{L'}{V'} t,\quad (x',y',z') = L' (x,y,z),\quad (u',v',w') = V' (u,v,w),\\
	\rho' &= \bar\rho' \rho, \qquad P' = \bar\rho' V'^2 P,\qquad T' = \frac{V'^2}{\mathfrak R'} T.\end{aligned}
\end{equation}
In this context, the typical scales are $L'=2\;\mathrm{km}$, $V'=20\;\mathrm{m\,s^{-1}}$ and $\bar\rho'=1\;\mathrm{kg\,m^{-3}}$, while $\mathfrak R'\approx 287\;\mathrm{m^2\,s^{-2}\,K^{-1}}$ is the gas constant for dry air, and, the non-dimensional constants in \eqref{eq:Euler2} and \eqref{eq:thermolaw} are $g = g'{L'}/{V'^2} \approx 49$ and  $\kappa = {\mathfrak R'}/{c_p'} \approx 0.287$,
where $g'$ is the gravitational acceleration and $c_p'\approx 1000\;\mathrm{m^2\,s^{-2}\,K^{-1}}$ is the specific heat of dry air at an atmospheric pressure of $1000\;\mathrm{mb}$. It is worth noting that, as the orographic lifting of air particles can be considered a dry adiabatic process outside regions of active precipitation (see \cite{20}), setting the right-hand side of \eqref{eq:thermolaw} to be zero is a reasonable assumption.\\
\\
The solution to the set of equations \eqref{eq:Euler2}, \eqref{mass}, \eqref{gas}, \eqref{eq:thermolaw} in \cite{ConstantinMountain2023} is obtained, in the Lagrangian setting, by specifying at any time $t$ the positions of the moving air particles
                \begin{equation} \label{solution}
                x  =\bar{u} t + a - \frac{1}{k}\,\mathrm{e}^{kb}\sin\theta,\qquad
                y=s,\qquad
                z =\bar{w} t + b + \frac{1}{k}\,\mathrm{e}^{kb}\cos\theta\,, \end{equation}
		in terms of the labelling variables $(a,s,b)\in(0,+\infty)\times(-s_0,s_0)\times (b_1,b_0)$ with $b_1<b_0<0$ and $s_0>0$, and parameters $k>0$ (wavenumber), $c=\sqrt{g/k}$ (wave speed), and $\bar{u}, \bar{w}  \in\boldsymbol{\mathbb{R}}$. Moreover,  we denote by    $\theta=k(a-ct)$.  See Fig. \ref{fig:path} and Fig. \ref{fig:path2}. Actually, in \cite{ConstantinMountain2023}, the author assumed \emph{a priori} that $v=0$ and that there is no $y$-dependence, that is, assuming the flow to be 2-dimensional.  However, for the purpose of our analysis, it is better to cast the 2-dimensional flow into a 3-dimensional framework.  Since in the derivation of the nonlinear system of governing equations \eqref{eq:Euler2}, \eqref{mass}, \eqref{gas}, \eqref{eq:thermolaw}, typical scales have been used for the non-dimensionalisation, it follows that $k=2\pi$ is the most realistic choice. Furthermore, it has to be noted that the restriction on the range of the label $b$ is a consequence of the fact that turbulence is usually present above and below the laminar layer governed by the equations  \eqref{eq:Euler2}, \eqref{mass}, \eqref{gas}, \eqref{eq:thermolaw}, and whose lower and upper boundaries correspond to $b=b_1$ and $b=b_0$, respectively. The constraint $b_0<0$ has to be imposed, since for $b\uparrow 0$, the vorticity of the flow (cf. equation \eqref{vorticity})        is unbounded. 
        
 \noindent 
 From \eqref{solution}, the velocity of a particle with labels  $(a,s,b)$ is obtained by computing the time derivative of its position, so that
        \begin{equation} \label{velocities} u =\bar{u} + c\,\mathrm{e}^{kb}\cos\theta,\qquad v=0,\qquad w=\bar{w} +c\,\mathrm{e}^{kb}\sin\theta \,. \end{equation}
      The Jacobian matrix associated with \eqref{solution} is  \begin{equation} \label{jacobian}\begin{pmatrix} x_a& x_s & x_b \\
     y_a & y_s& y_b \\ z_a & z_s & z_b \end{pmatrix} = \begin{pmatrix} 1-\mathrm{e}^{kb}\cos \theta  & 0& -\,\mathrm{e}^{kb}\sin \theta \\
        0 & 1 & 0\\-\,\mathrm{e}^{kb}\sin \theta \qquad &0& 1+\mathrm{e}^{kb}\cos \theta \end{pmatrix}\, \end{equation}
         with determinant and inverse given, respectively, by
                \begin{equation} \D(b)=1-\mathrm{e}^{2kb}>0
                 \end{equation}
                and
                 \begin{equation}\label{jacob}\begin{pmatrix}
a_x & a_y & a_z \\
s_x & s_y & s_z \\
b_x & b_y & b_z
\end{pmatrix}=\frac{1}{1-e^{2kb}}
\begin{pmatrix}
1+e^{kb}\cos\theta& 0 & e^{kb}\sin\theta\\
0 & 1-e^{2kb} & 0\\
e^{kb}\sin\theta & 0 & 1-e^{kb}\cos\theta
\end{pmatrix}.
\end{equation}
The vorticity of the flow is 
\begin{equation}\label{vorticity}
            \boldsymbol{\omega}=(\omega_1,\omega_2,\omega_3)=\left(0,\,-\dfrac{2kc\, e^{2kb}}{1 -e^{2kb}},\,0\right).
        \end{equation}
As the determinant is time-independent and positive, it follows that (see \cite{C_Book})\begin{equation}\label{div free}
    u_x+v_y+w_z=0,
\end{equation}  and, for a density of the form $\rho=\rho(b)$ (decreasing with $b$) as in \cite{ConstantinMountain2023}, it holds that $\rho_t + u\,\rho_x + v\,\rho_y + w\,\rho_z=0$, ensuring that \eqref{mass} is satisfied. Lastly, with a pressure distribution of the form
\begin{equation}
    P=P^* + g \int_{b_1}^b \rho(\tilde{b})(e^{2k\tilde{b}}-1)d\tilde{b}, 
\end{equation}
with $b_1<b_0<0$ and $P^*>0$, the velocity field \eqref{velocities} solves the system given by \eqref{eq:Euler2} and \eqref{mass}. Finally, given the solution $(u,v,w,P,\rho)$, the associated temperature $T$ can be determined from \eqref{gas}, under the compatibility condition \eqref{eq:thermolaw} (see \cite{ConstantinMountain2023} for more details).\\
Before proceeding further, let us derive some other standard thermodynamic relations, which were not present in \cite{ConstantinMountain2023}, but will prove very useful in the subsequent instability analysis. Combining \eqref{mass}, \eqref{gas} and \eqref{eq:thermolaw}, we obtain
\begin{equation}
    \frac{D P}{Dt} + \frac{1}{1-\kappa} P \nabla\cdot \bu =0\qquad\text{and}\qquad
\frac{D\ln T}{Dt}=\frac{\kappa}{1-\kappa}\frac{D\ln\rho}{Dt},
\end{equation}
so that dividing by $T$ and integrating along particle paths, yields $
T \propto \rho^{\kappa/(1-\kappa)}$, leading to
$P=\rho T \propto \rho^{\,1+\kappa/(1-\kappa)}=\rho^{\,1/(1-\kappa)}$, resulting in the isentropic relation for ideal gases:
\begin{equation}\label{isentropic}
   P=K\,\rho^\gamma, \qquad\text{where} \qquad \gamma=\frac{1}{1-\kappa},
\end{equation}
and $K$ is constant along particle paths (see e.g. \cite{cengel}). From \eqref{div free} it also follows that ${DP}/{Dt}=0.$

\section{Short-wavelength instability analysis for trochoidal compressible flows}\label{Sec:instability}
In order to prove that the above solution is (linearly) unstable, we adopt the short-wavelength instability method of \cite{LH} (and further developed in \cite{LebovitzL}) for compressible flow. We consider small perturbations of the velocity, pressure, and density fields, denoted by
\[
\boldsymbol{\mathfrak{u}}
=(
\mathfrak{u},\,
\mathfrak{v},\,
\mathfrak{w}),
\qquad
\mathfrak{p},
\qquad
\varrho,
\]
respectively. Thus, the perturbed fields are written as
\begin{equation}
    \bu \rightarrow \bu+\boldsymbol{\mathfrak{u}},
    \qquad
    P\rightarrow P+\mathfrak{p},
    \qquad
    \rho\rightarrow \rho+\varrho.
\end{equation}
The perturbations are superimposed on the background flow $(\bu,P,\rho)$ which is 
governed by the general equations of fluid motion
\begin{equation}
    \frac{D\bu}{Dt}
    +\frac{\nabla P}{\rho}
    +\begin{pmatrix}
        0\\
        0\\
        g
    \end{pmatrix}
    =0,
    \qquad
    \frac{D\rho}{Dt}
    +\rho\nabla\cdot\bu
    =0,
    \qquad
    \frac{DP}{Dt}
    +\gamma P\nabla\cdot\bu
    =0.
\end{equation}

Substituting the perturbed fields into these equations and retaining only
terms that are linear in the perturbations yields the linearised equations
governing their evolution. In particular, using the expansion
\[
\frac{1}{\rho+\varrho}
=
\frac{1}{\rho}
-\frac{\varrho}{\rho^2}
+\mathcal{O}(\varrho^2),
\]
and neglecting all terms of quadratic or higher order in
$\boldsymbol{\mathfrak{u}}$, $\mathfrak{p}$, and $\varrho$, we obtain 
\begin{equation}\label{lin perturbations}
    \left\{\begin{aligned}
   & \frac{ D\uu}{Dt}+\uu\cdot\nabla \bu +\frac{\nabla \mathfrak{p}}{\rho}-\varrho\frac{\nabla P}{\rho^2}=0,\\
   & \frac{D\varrho}{Dt}+\boldsymbol{\mathfrak{u}} \cdot\nabla\rho+\rho \nabla\cdot\boldsymbol{\mathfrak{u}}+\varrho\nabla\cdot\bu=0,\\
    &\frac{D\mathfrak{p}}{Dt}+\boldsymbol{\mathfrak{u}} \cdot\nabla P+\gamma P \nabla\cdot\boldsymbol{\mathfrak{u}}+\gamma\mathfrak{p}\nabla\cdot\bu=0,
\end{aligned}\right.
\end{equation}
where ${D}/{Dt}={\partial}/{\partial t} + \bu\cdot\nabla$ is the material derivative along the basic flow $\bu$, and $\nabla\bu$ is the velocity-gradient tensor. The initial conditions for the system \eqref{lin perturbations} are
\begin{equation}
    \uu(\x,0)=\uu_0(\x),\qquad  \mathfrak{p}(\x,0)=\mathfrak{p}_0(\x),\qquad  \varrho(\x,0)=\varrho_0(\x).
\end{equation}
Recalling that $\nabla\cdot\bu=0$ (see equation \eqref{div free}), the second and third equations in \eqref{lin perturbations} can be reduced to 
\begin{equation}\label{2 and 3}\frac{D\varrho}{Dt}+\boldsymbol{\mathfrak{u}} \cdot\nabla\rho+\rho \nabla\cdot\boldsymbol{\mathfrak{u}}=0,\qquad \text{and} \qquad
    \frac{D\mathfrak{p}}{Dt}+\boldsymbol{\mathfrak{u}} \cdot\nabla P+\gamma P \nabla\cdot\boldsymbol{\mathfrak{u}}=0,
\end{equation}
respectively. Following  \cite{LH, LebovitzL}, we adopt the Eckart transform (see also\cite{eckart}) to introduce new variables $\m$ and $\nn$ such that
\begin{equation}\label{eckart variables}
    \varrho=\frac{\rho}{\cs}(\m+\nn) \qquad\text{and}\qquad \mathfrak{p}=\frac{\gamma P}{\cs}\nn=\rho\cs\nn,
\end{equation}
where $\cs$ is the local speed of sound defined as $\cs=\sqrt{\gamma {P}/{\rho}}$, and $\gamma=({1-\kappa})^{-1}$. Substituting the new variables \eqref{eckart variables} into the system of equations \eqref{lin perturbations} and \eqref{2 and 3}, and using the fact that ${D\rho}/{Dt}={DP}/{Dt}={D\cs}/{Dt}=0$, we obtain the linearized equation of motion for the perturbations in the new variables, namely
\begin{equation}\left \{
\begin{aligned} 
  & \frac{ D\uu}{Dt}+\uu\cdot\nabla \bu-\mathfrak{m}\cs\nabla g_1 + \cs\nabla\mathfrak{n}+\mathfrak{n}\cs\nabla g_2=0,\\
  & \frac{D \m}{Dt} +\uu\cdot\cs\nabla g_3=0,\\
  & \frac{D \nn}{Dt}+\cs\nabla\cdot\uu+\uu\cdot\cs\nabla g_1=0,\label{45c}
   \end{aligned}\right.
\end{equation}  
where 
\begin{equation}\label{g's}
    g_1 =\ln\left(P^{\frac{1}{\gamma}}\right)\qquad g_2=\ln\left(\frac{\gamma P^{\frac{\gamma-1}{\gamma}}}{\cs}\right)\qquad g_3 = \ln\left(\frac{\rho}{P^{\frac{1}{\gamma}}}\right).
\end{equation}
It is important to note that the functions $g_1$, $g_2$ and $g_3$ in \eqref{g's} are defined solely in terms of the basic flow variables and do not depend on the perturbations. 

\noindent
The short-wavelength instability approach consists in investigating the boundedness in time of the amplitude of rapidly oscillating, localized perturbations evolving according to \eqref{45c}, whose initial data have the form
\begin{equation}
    \bigl[\uu_0(\x), \m_0(\x), \nn_0(\x)\bigr]= \bigl[\hat{\uu}_0(\x), \hat{\m}_0(\x), \hat{\nn}_0(\x)\bigr]\, e^{{\frac{i }{\e}\Phi_0(\x)}}
\end{equation}
with $\e$ a small parameter, and $\hat{\uu}_0$ is chosen as follows
\begin{equation}
    \hat{\uu}_0(\x)=P^{-\frac{1}{\gamma}}\left(\nabla F(\x)\right)\times \left(\nabla \Phi_0(\x)\right),
\end{equation}
where $F$ is a smooth function, so that 
\begin{equation}
    \nabla\cdot \left(\hat{\uu}_0\,P^{\frac{1}{\gamma}}\right)=0, \qquad\text{and}\qquad \left(\hat{\uu}_0 \,P^{\frac{1}{\gamma}}\right)\cdot\nabla\Phi_0=0.
\end{equation}
It can be proved (see \cite{LebovitzL}) that the solution is given as 
\begin{equation}\label{up}
    \begin{aligned}
      \left[  \boldsymbol{\mathfrak{u}}(\x,t),\, \m(\x,t),\, \nn(\x,t)\right]=\left[\A(\x,t),\, \B(\x,t),\,0\right] e^{\frac{i}{\e}\Phi(\x,t)}+\mathcal{O}(\e)
    \end{aligned}
\end{equation}
where $\A=(\mathfrak{U}_1, \mathfrak{U}_2, \mathfrak{U}_3)$ is a vector function, while $\B$ and $\Phi$ are scalar functions. Their evolution, at leading order in $\e$, is given by the following coupled system of ODEs 
  \begin{equation}\label{leading order}
   \left \{\begin{aligned}
    \dot{\x}&=\bu(\x,t),\\
     \dot{\boldsymbol{\xi}}& =-(\nabla\bu)^T\boldsymbol{\xi}, \\  
    \dot{\A}&=-(\A\cdot\nabla)\bu+2\dfrac{\boldsymbol{\xi}\cdot[(\A\cdot\nabla)\bu]}{||\boldsymbol{\xi}||^2}\boldsymbol{\xi}+\cs\left(\nabla g_1-\dfrac{\nabla g_1\cdot\boldsymbol{\xi}}{||\boldsymbol{\xi}||^2}\boldsymbol{\xi}\right)\B, \\
       \dot{\B}&=-\A\cdot\cs \nabla g_3 ,
       \end{aligned} \right.
\end{equation}
In \eqref{leading order}, we have introduced $\boldsymbol{\xi}:=\nabla\Phi$, and denoted by $\dot{\empty}=D/{Dt}$. The system \eqref{leading order} is coupled with the initial conditions
\begin{equation}
    \x(0)=\x_0,\qquad \boldsymbol{\xi}(0)=\boldsymbol{\xi}_0, \qquad \A(0)=\A_0=\hat{\uu}_0(\x_0), \qquad \B(0)=\B_0={\hat{{\m}}_0(\x_0)},
\end{equation}
satisfying the constraint \(
    \A_0\cdot \boldsymbol{\xi}_0=0.\) Lastly, we also recall that the function $\Phi$ satisfies the eikonal equation $D\Phi/Dt=0$, with initial data $\Phi(\x,0)=\Phi_0(\x)$ (see \cite{LebovitzL}). The first equation in \eqref{leading order} describes the particle trajectory of the basic flow (see \cite{Bennett}), while the other three describe---at leading order in $\e$---the evolution of the local wave vector and the amplitude of the perturbation along the particle trajectory, respectively. This system therefore describes the evolution in time of a sharply-peaked initial disturbance, located at time $t=0$ around $ \x_0$, that is advected by the basic flow $ \bu$, and in doing so, its amplitude may or may not increase. If it increases, that is
    \begin{equation}
        \sup_{\substack{\x_0,\boldsymbol{\xi}_0, \A_0,\B_0\\ \text{s.t.}\ \A_0\cdot \boldsymbol{\xi}_0=0 }} \limsup_{t\rightarrow\infty}\bigl|\left[\A(\x_0,\boldsymbol\ {\xi}_0, \A_0,\B_0;\,t),\, \B(\x_0,\boldsymbol{\xi}_0, \A_0,\B_0;\,t)\right]\bigr|=\infty
    \end{equation}the flow is unstable (see \cite{LH,LebovitzL} for more details, or \cite{Kruse2} for a review of this method in the case of incompressible geophysical flows). \\
    \\
    With the intention of applying this method to the proposed solution of \cite{ConstantinMountain2023}, we compute the gradient tensor of the basic flow $\bu$ to get    
\begin{equation}\label{nablaU}
\begin{aligned}
 \nabla \bu&=\begin{pmatrix}
	u_x & u_y & u_z\\
	v_x & v_y & v_z\\
	w_x & w_y & w_z
\end{pmatrix} = \begin{pmatrix}
	u_a & u_s & u_b\\
	v_a & v_s & v_b\\
	w_a & w_s & w_b
\end{pmatrix}  \begin{pmatrix}
	a_x & s_x & b_x \\
	a_y & s_y & b_y\\
	a_z& s_z & b_z		
\end{pmatrix}\\
&=\dfrac{1}{\D}\begin{pmatrix}
	 -kce^{kb}\sin\theta & 0& kce^{kb}(\cos\theta-e^{kb})\\
     0 &0&0\\
  kc e^{kb}(\cos\theta + e^{kb})& 0 & kc e^{kb}\sin\theta
\end{pmatrix}.
\end{aligned}
\end{equation}
Thus, if $\boldsymbol{\xi}_0=(0,1,0)^T$, from the second of \eqref{leading order} it follows that $\dot{\boldsymbol{\xi}}=0$, giving $\boldsymbol{\xi}(t)=(0,1,0)^T$ for all $t$. Moreover, note that, due to \eqref{nablaU}, we have $\boldsymbol{\xi}\cdot[(\A\cdot\nabla)\bu]=0$, as well as $\nabla g_1\cdot\boldsymbol{\xi}=0$ since \eqref{eq:Euler2} coupled with \eqref{velocities} ensures that $P_y=0$. Furthermore, due to \eqref{jacob}, it follows that $\rho_y=\rho_b b_y =0$.\\
Therefore, the last two equations of \eqref{leading order} give
\begin{equation} \label{ode}
    \left\{\begin{aligned}
        \begin{pmatrix}
            \dot{\mathfrak{U}_1}\\
            \dot{\mathfrak{U}_2}\\
            \dot{\mathfrak{U}_3}
        \end{pmatrix}&=\begin{pmatrix}
	 \dfrac{kce^{kb}\sin\theta}{1-e^{2kb}} & 0& \dfrac{kce^{kb}(e^{kb}-\cos\theta)}{1-e^{2kb}}\\
     0 &0&0\\
    -\dfrac{kc e^{kb}(\cos\theta + e^{kb})}{1-e^{2kb}}& 0 & -\dfrac{kc e^{kb}\sin\theta}{1-e^{2kb}}
\end{pmatrix}\begin{pmatrix}
            {\mathfrak{U}_1}\\
            {\mathfrak{U}_2}\\
            {\mathfrak{U}_3}
        \end{pmatrix}       +\dfrac{1}{\cs\rho}\begin{pmatrix}
           P_x\\
            0\\
            P_z
        \end{pmatrix}\B, \\ \vspace{.5cm}
        \dot{\B}&=-\cs({\mathfrak{U}_1}\, g_{3,\,x}+{\mathfrak{U}_2}\, g_{3,\,y}+{\mathfrak{U}_3}\, g_{3,\,z}).
    \end{aligned}\right.
\end{equation}
In particular $\dot{\mathfrak{U}_2}=0$ and the constraint \(    \A_0\cdot \boldsymbol{\xi}_0=0\) forces $\mathfrak{U}_2(0)=0$, giving $\mathfrak{U}_2(t)=0$ for all $t$. The system \eqref{ode} therefore reduces to the following $3\times3$ linear system 
\begin{equation}\label{3ode}
\begin{pmatrix}
            \dot{\mathfrak{U}_1}\\
            \dot{\mathfrak{U}_3}\\
            \dot{\mathfrak{M}}
        \end{pmatrix}
=\underbrace{
\begin{pmatrix}
\dfrac{kce^{kb}\sin\theta}{1-e^{2kb}}
&
\dfrac{kce^{kb}(e^{kb}-\cos\theta)}{1-e^{2kb}}
&
\dfrac{P_x}{\cs\rho}
\\[1.2em]
-\dfrac{kc e^{kb}(\cos\theta+e^{kb})}{1-e^{2kb}}
&
-\dfrac{kce^{kb}\sin\theta}{1-e^{2kb}}
&
\dfrac{P_z}{\cs\rho}
\\[1.2em]
-\cs g_{3,\,x}
&
-\cs g_{3,\,z}
&
0
\end{pmatrix}}_{:=\mathbb{B}}\begin{pmatrix}
            {\mathfrak{U}_1}\\
            {\mathfrak{U}_3}\\
            {\mathfrak{M}}
        \end{pmatrix}
\end{equation}
Recalling that $g_3 = \ln\left({\rho}/{P^{\frac{1}{\gamma}}}\right)$, and that $\rho$ and $P$ (and consequently $\cs$) depend solely on the labeling variable $b$, we have $g_{3,\,x}=b_x  g_{3,\,b}$ and $g_{3,\,z}=b_z  g_{3,\,b}$, with
\begin{equation}
\begin{aligned}
    b_x&=\frac{e^{kb}\sin\theta}{1-e^{2kb}}, \qquad b_z=\frac{1-e^{kb}\cos\theta}{1-e^{2kb}},\\
     g_{3,\,b}&=\frac{\rho_b}{\rho}-\frac{P_b}{\gamma P}=\frac{\rho_b}{\rho}+\frac{g}{\cs^2}(1-e^{2kb}),\\
P_b&=g\rho(e^{2kb}-1),
\end{aligned}
\end{equation}
so that the matrix in \eqref{3ode} reads as
\begin{equation}\label{B matrix}
  \mathbb{B}=  \begin{pmatrix}
\dfrac{kc\ee\sin\theta}{\D}
&
\dfrac{kc\ee(\ee-\cos\theta)}{\D}
&
-\dfrac{g}{\cs}{\ee\sin\theta}
\\[1.2em]
-\dfrac{kc \ee(\cos\theta+\ee)}{\D}
&
-\dfrac{kc\ee\sin\theta}{\D}
&
-\dfrac{g}{\cs}({1-\ee\cos\theta})
\\[1.2em]
-\cs\dfrac{\ee\sin\theta}{\D}\left(\dfrac{\rho_b}{\rho}+\dfrac{g\D}{\cs^2}\right)
&
-\cs \dfrac{1-\ee\cos\theta}{\D}\left(\dfrac{\rho_b}{\rho}+\dfrac{g\D}{\cs^2}\right)
&
0
\end{pmatrix},
\end{equation}
where
\begin{equation}
    (0,1)\ni\ee:=e^{kb}, \qquad (0,1)\ni\D:=1-e^{2kb}.
\end{equation}
Unfortunately, this system cannot be put---in general---into an autonomous form by a simple rotation, as is the case for incompressible flows (see, e.g., \cite{CG, Kruse, Kruse2, mioInstability}). Nevertheless, the coefficients are periodic, which allows us to characterize the instability by means of Floquet theory. We therefore adopt a Floquet-theoretic approach to analyze the stability properties of the system, with the aim of identifying the parameter regimes in which perturbations grow and of characterizing the corresponding instability mechanisms.


\subsection{Floquet analysis}\label{sec:floquet}

\subsubsection{Reduction to a two-parameter periodic problem}
Along a particle path the labels $(a,s,b)$ are constant. Hence $\ee=e^{kb}$, $\D=1-\ee^2$, $\rho$, $\rho_b$, $P$ and $\cs$ are constant, and $\mathbb B$ depends on time only through $\theta=k(a-ct)$, for which $\dot\theta=-kc$. Before proceeding further, let us note that $c^2=g/k$ gives $g/(kc)=c$. Moreover, we rescale the entropy amplitude as
\begin{equation}\label{eq:rescale}
\widetilde{\B}:=\frac{c}{\cs}\,\B,
\end{equation}
and introduce the following stratification parameter,
\begin{equation}\label{eq:nu}
\nu^2:=-\frac{g_{3,b}}{k}=-\frac1k\left(\frac{\rho_b}{\rho}+\frac{g\D}{\cs^2}\right)
=-\frac{\rho_b}{k\rho}-\frac{c^2}{\cs^2}\,\bigl(1-\ee^2\bigr).
\end{equation}
Since $\gamma^{-1}=1-\kappa$, the non-dimensional potential temperature $\Theta=T P^{-\kappa}$ satisfies $\ln\Theta=\gamma^{-1}\ln P-\ln\rho$, so that $g_3=-\ln\Theta$ up to an additive constant. Consequently
\begin{equation}\label{eq:nuphys}
\nu^2=\frac{1}{k}\frac{{\rm d}\ln\Theta}{{\rm d}b}=\frac{\mathcal N^2}{\Omega^2},\qquad\text{where}\qquad
\mathcal N^2:=g\,\frac{{\rm d}\ln\Theta}{{\rm d}b},\qquad \Omega:=kc=\sqrt{gk},
\end{equation}
i.e.\ $\nu$ is the ratio between the Brunt--V\"ais\"al\"a frequency measured in the label coordinate $b$ (the Eulerian one being $N^2=\mathcal N^2 b_z$) and the angular frequency $\Omega$ of the orbital motion of the air parcels. The sign of $\nu^2$ records the stratification:
\begin{itemize}
\item $\nu^2>0$: stable (statically stable) stratification;
\item $\nu^2=0$: neutral stratification -- a well-mixed, dry-adiabatic layer, $K=$ const in \eqref{isentropic};
\item $\nu^2<0$: statically unstable stratification.
\end{itemize}
With $\boldsymbol{\mathcal E}=(\mathfrak{U}_1,\mathfrak{U}_3,\widetilde{\B})^T$ the system becomes
\begin{equation}\label{eq:Asys}
\boldsymbol{\mathcal E}'=\mathbb A(\theta)\boldsymbol{\mathcal E},\qquad\text{with}\qquad
\mathbb A(\theta)=
\begin{pmatrix}
-\dfrac{\ee\sin\theta}{\D} & \dfrac{\ee(\cos\theta-\ee)}{\D} & \ee\sin\theta\\[1.2ex]
\dfrac{\ee(\cos\theta+\ee)}{\D} & \dfrac{\ee\sin\theta}{\D} & 1-\ee\cos\theta\\[1.2ex]
-\nu^2\dfrac{\ee\sin\theta}{\D} & -\nu^2\dfrac{1-\ee\cos\theta}{\D} & 0
\end{pmatrix},
\end{equation}
and we have denoted by $'={\rm d}/{\rm d}\theta$.

\begin{remark}\label{rem:params}
The uniform velocities $\bar{u},\bar{w}$ do not appear in \eqref{eq:Asys}, and the Mach number $c/\cs$ enters only through $\nu^2$. For a given level $b$ the stability problem therefore depends on two numbers only: the steepness $\ee=e^{kb}\in(0,1)$ of the trochoidal isoline $b=\mathrm{const}$ (whose amplitude is $e^{kb}/k$), and the stratification parameter $\nu^2$.
\end{remark}
\noindent
It is convenient to write $\mathbf V=(\mathfrak{U}_1,\mathfrak{U}_3)^T$ and to introduce the $2\pi$-periodic vectors
\begin{equation}\label{eq:pq}
\mathbf p(\theta)=\begin{pmatrix}1-\ee\cos\theta\\-\ee\sin\theta\end{pmatrix}=\begin{pmatrix}
    x_a\\ z_a
\end{pmatrix},\qquad
\mathbf q(\theta)=\begin{pmatrix}\ee\sin\theta\\1-\ee\cos\theta\end{pmatrix}=J\mathbf p=\D\,\begin{pmatrix}
    b_x\\ b_z
\end{pmatrix},
\end{equation}
where $\displaystyle J=\begin{pmatrix}0&-1\\1&0\end{pmatrix}$. The vectors $\mathbf p$ and $\mathbf q$ are tangent and normal, respectively, to the isolines $b=\mathrm{const}$. They satisfy $\mathbf p\cdot\mathbf q=0$ and $|\mathbf p|^2=|\mathbf q|^2=h(\theta):=1-2\ee\cos\theta+\ee^2>0$. Then \eqref{eq:Asys} reads
\begin{equation}\label{eq:compact}
\begin{aligned}
    \mathbf V'&=\Msf\,\mathbf V+\widetilde{\B}\,\mathbf q,\\
\widetilde{\B}'&=-\frac{\nu^2}{\D}\,\mathbf q\cdot\mathbf V,
\end{aligned}\qquad
\Msf:=\frac{1}{kc}\begin{pmatrix}u_x&u_z\\w_x&w_z\end{pmatrix}
=\frac{\ee}{\D}\begin{pmatrix}-\sin\theta&\cos\theta-\ee\\ \cos\theta+\ee&\sin\theta\end{pmatrix}.
\end{equation}
Let us note that
\begin{equation}\label{eq:ptransport}
\mathbf p'=-\Msf\,\mathbf p.
\end{equation}
This can be checked directly: since $\mathbf p'=\ee\,(\sin\theta,\,-\cos\theta)^T$,
\begin{equation}
\Msf\mathbf p=\frac{\ee}{\D}\begin{pmatrix}-\sin\theta(1-\ee\cos\theta)-\ee\sin\theta(\cos\theta-\ee)\\ (\cos\theta+\ee)(1-\ee\cos\theta)-\ee\sin^2\theta\end{pmatrix}
=\ee\begin{pmatrix}-\sin\theta\\ \cos\theta\end{pmatrix}=-\mathbf p'.
\end{equation}
Moreover, the antisymmetric part of $\Msf$ is related to the vorticity \eqref{vorticity} by
\begin{equation}\label{eq:antisym}
\Msf-\Msf^T=\frac{2\ee^2}{\D}\,J=-\frac{\omega_2}{kc}\,J,\qquad \omega_2=u_z-w_x=-\frac{2kc\,\ee^{2}}{\D}.
\end{equation}
Before proceeding with the Floquet analysis, let us note that the system \eqref{eq:compact} possesses the following conserved quantity.

\begin{proposition}[Conserved quantity]\label{prop:PV}
The scalar
\begin{equation}\label{eq:I}
\mathcal I:=\nu^2\,\mathbf p\cdot\mathbf V-2\ee^2\,\widetilde{\B}
\end{equation}
is constant along every solution of \eqref{eq:compact}.
\end{proposition}
\begin{proof}
By \eqref{eq:ptransport} and \eqref{eq:compact},
\[
(\mathbf p\cdot\mathbf V)'=\mathbf p'\cdot\mathbf V+\mathbf p\cdot\mathbf V'
=-(\Msf\mathbf p)\cdot\mathbf V+\mathbf p\cdot(\Msf\mathbf V)+\Big(\widetilde{\B}\,\underbrace{\mathbf p\cdot\mathbf q}_{=0}\Big).
\]
Since $(\Msf\mathbf p)\cdot\mathbf V=\mathbf p\cdot(\Msf^T\mathbf V)$ for any matrix $\Msf$, the two remaining terms combine into
\[
(\mathbf p\cdot\mathbf V)'=\mathbf p\cdot\big[(\Msf-\Msf^T)\mathbf V\big].
\]
By \eqref{eq:antisym}, $\Msf-\Msf^T=\frac{2\ee^2}{\D}J$; and since $J^T=-J$ and $J\mathbf p=\mathbf q$,
\[
\mathbf p\cdot(J\mathbf V)=(J^T\mathbf p)\cdot\mathbf V=-(J\mathbf p)\cdot\mathbf V=-\mathbf q\cdot\mathbf V,
\]
so that
\begin{equation}\label{eq:pVprime}
(\mathbf p\cdot\mathbf V)'=\frac{2\ee^2}{\D}\,\mathbf p\cdot(J\mathbf V)=-\frac{2\ee^2}{\D}\,\mathbf q\cdot\mathbf V.
\end{equation}
On the other hand, the second ODE in \eqref{eq:compact} gives directly
\begin{equation}\label{eq:MtildeVprime}
\big(2\ee^2\widetilde{\B}\big)'=2\ee^2\widetilde{\B}'=-\frac{2\ee^2\nu^2}{\D}\,\mathbf q\cdot\mathbf V.
\end{equation}
Combining \eqref{eq:pVprime} and \eqref{eq:MtildeVprime} in the definition \eqref{eq:I} of $\mathcal I$,
\[
\mathcal I'=\nu^2(\mathbf p\cdot\mathbf V)'-\big(2\ee^2\widetilde{\B}\big)'
=\nu^2\Big(-\frac{2\ee^2}{\D}\,\mathbf q\cdot\mathbf V\Big)-\Big(-\frac{2\ee^2\nu^2}{\D}\,\mathbf q\cdot\mathbf V\Big)=0. 
\]
\end{proof}
\noindent
The conserved quantity $\mathcal I$ is the leading-order contribution to the Ertel potential vorticity of the rapidly oscillating perturbations; see Appendix~\ref{app:PV}.

\subsubsection{Basics of Floquet theory}
Since the system \eqref{eq:Asys} (and consequently \eqref{eq:compact}) has periodic coefficients, its analysis naturally falls within the framework of Floquet theory; see, e.g., \cite{YakubovichStarzhinskii1975}. We briefly recall the relevant basic facts.\\
\noindent
Let $\F(\theta)$ denote the fundamental matrix solution of the $3\times3$ linear system \eqref{eq:Asys}, normalised by\begin{equation}
\F'(\theta)=\mathbb A(\theta)\F(\theta),
\qquad
\F(0)=\mathbb I.
\end{equation}
Thus, the columns of $\F(\theta)$ form a basis of linearly independent solutions of the system, and, for any initial condition $\boldsymbol{\mathcal{E}}(0)$ the corresponding solution of \eqref{eq:Asys} is given by
\begin{equation}
\boldsymbol{\mathcal{E}}(\theta)=\F(\theta)\boldsymbol{\mathcal{E}}(0)
\end{equation}
Since $\mathbb A(\theta+2\pi)=\mathbb A(\theta)$, both $\F(\theta+2\pi)$ and $\F(\theta)\F(2\pi)$ solve the same initial-value problem. Hence, by uniqueness,
\begin{equation}\label{eq:period}
\F(\theta+2\pi)=\F(\theta)\F(2\pi),
\qquad
\F(2\pi n)=\F(2\pi)^n,
\quad n\in\mathbb Z.
\end{equation}
The matrix
\[
\mathbb M:=\F(2\pi)
\]
is the {monodromy matrix}, and its eigenvalues $\mu_1,\mu_2,\mu_3$ are the {Floquet multipliers}. In particular, if $\boldsymbol{\mathcal E}(0)$ is an eigenvector of $\mathbb M$ associated with the multiplier $\mu$, then
\[
\boldsymbol{\mathcal E}(2\pi n)
=\mathbb M^n\boldsymbol{\mathcal E}(0)
=\mu^n\boldsymbol{\mathcal E}(0).
\]
Thus, over successive orbital periods, the corresponding mode is amplified or attenuated according to the factor $\mu^n$: a multiplier with $|\mu|>1$ gives exponential growth, whereas $|\mu|<1$ gives exponential decay. Equivalently, introducing the {Floquet exponent} $\zeta$ through $\mu=e^{2\pi\zeta}$, the exponential growth rate per unit $\theta$ is
\begin{equation}\label{eq:rate}
\operatorname{Re}\zeta
=\frac{\ln|\mu|}{2\pi},
\end{equation}
and, since $|\dot\theta|=\Omega$, the corresponding growth rate per unit physical time is $\sigma=\Omega\operatorname{Re}\zeta$. Consequently, the existence of a Floquet multiplier with $|\mu|>1$ is equivalent to the presence of an exponentially growing solution.

\begin{remark}[From $\theta$ to physical time]\label{rem:time}
Physical time increases as $\theta$ {decreases}, since $\dot\theta=-kc<0$. Consider the particle with label $a$, for which $\theta(t)=ka-\Omega t$ and $\boldsymbol{\mathcal E}(t)=\F(\theta(t))\F(ka)^{-1}\boldsymbol{\mathcal E}(0)$. A time step $t\mapsto t+T$, with $T=2\pi/\Omega$, corresponds to $\theta\mapsto\theta-2\pi$, and $\F(\theta-2\pi)=\F(\theta)\mathbb M^{-1}$ yields
\begin{equation}\label{eq:oneperiodmap}
\boldsymbol{\mathcal E}(t+T)=\mathbb M_t\,\boldsymbol{\mathcal E}(t),\qquad
\mathbb M_t:=\F\bigl(\theta(t)\bigr)\,\mathbb M^{-1}\,\F\bigl(\theta(t)\bigr)^{-1}.
\end{equation}
The one-period map in physical time is therefore similar to $\mathbb M^{-1}$, for every instant $t$ and every label $a$; its multipliers are the reciprocals $1/\mu_i$. As shown in Proposition~\ref{prop:factor} below, the multipliers of $\mathbb M$ are of the form $\{1,\mu,\mu^{-1}\}$, a set invariant under $\mu\mapsto\mu^{-1}$; hence $\mathbb M_t$ and $\mathbb M$ have the same multipliers and the direction in which $\theta$ is traversed is immaterial for the stability conclusions.
\end{remark}

\subsubsection{Structure of the monodromy matrix}
We now collect some  structural properties of $\mathbb M=\F(2\pi)$.

\begin{lemma}[Volume conservation]\label{lem:det}
$\det\F(\theta)\equiv1$ for all $\theta$; hence $\mu_1\mu_2\mu_3=1$.
\end{lemma}
\begin{proof}
We have $\tr\mathbb A=-\ee\sin\theta/\D+\ee\sin\theta/\D+0=0$. By Liouville's formula,
\begin{equation}
\det\F(\theta)=\exp\!\int_0^\theta\tr\mathbb A(s)\,{\rm d}s=1 \qquad\text{for all }\theta .
\end{equation}
Therefore $\det\mathbb M=\det\F(2\pi)=1$, and the three Floquet multipliers $\mu_1,\mu_2,\mu_3$ obey
\begin{equation}
\mu_1\mu_2\mu_3=1. 
\end{equation}
\end{proof}
\noindent
Physically, $\tr\mathbb A=0$ reflects the incompressibility of the basic flow, $\nabla\cdot\bu=0$ (eq.~\eqref{div free}): the linearised WKB flow map inherits this volume preservation.

\begin{lemma}\label{lem:rev}
Let $\s=\mathrm{diag}(1,-1,1)$. Then $\s\,\mathbb A(-\theta)\,\s=-\mathbb A(\theta)$ for all $\theta$, and consequently
\begin{equation}\label{eq:rev}
\mathbb M=\s\,\mathbb M^{-1}\,\s.
\end{equation}
\end{lemma}
\begin{proof}
Conjugation by $\s$ multiplies the $(i,j)$ entry of a $3\times3$ matrix according to the sign pattern $\left(\begin{smallmatrix}+&-&+\\-&+&-\\+&-&+\end{smallmatrix}\right)$. Inspection of $\mathbb A(\theta)$ shows that the entries proportional to $\sin\theta$ occupy exactly the ``$+$'' positions, and those proportional to $\cos\theta$ or constant occupy exactly the ``$-$'' positions. Since $\sin(-\theta)=-\sin\theta$ while $\cos(-\theta)=\cos\theta$, replacing $\theta\to-\theta$ already flips the sign of the ``$+$''-position entries; conjugating by $S$ then flips the sign of the remaining entries as well. Combining the two sign flips gives $\s\,\mathbb A(-\theta)\,\s=-\mathbb A(\theta)$.

Now set $\Psi(\theta):=\s\,\F(-\theta)\,\s$. Then $\Psi(0)=\mathbb I$ and
\[
\Psi'(\theta)=-\s\,\mathbb A(-\theta)\,\F(-\theta)\,\s=-\big[\s\,\mathbb A(-\theta)\,\s\big]\big[\s\,\F(-\theta)\,\s\big]=\mathbb A(\theta)\,\Psi(\theta),
\]
using $\s^2=\mathbb I$. By uniqueness, $\Psi\equiv\F$, i.e.\ $\F(\theta)=\s\,\F(-\theta)\,S$ for all $\theta$. Taking $\theta=-2\pi$ in \eqref{eq:period} gives $\F(-2\pi)=\mathbb M^{-1}$; combined with $\F(-2\pi)=\s\,\F(2\pi)\,\s$ (the identity just proved, at $\theta=2\pi$), this yields \eqref{eq:rev}.
\end{proof}
\noindent
Equation \eqref{eq:rev} ensures that $\mathbb M$ is similar to its own inverse, so its spectrum is invariant under $\mu\mapsto1/\mu$. 

\begin{proposition}[Factorisation of the characteristic polynomial]\label{prop:factor}
Let $p:=\tr\mathbb M$. Then one of the Floquet multipliers equals
\begin{equation}\label{eq:mu1}
\mu_1=1,
\end{equation}
irrespective of the values of the parameters, while the remaining two are the roots of
\begin{equation}\label{eq:mu23eq}
\mu^{2}-(p-1)\,\mu+1=0,
\end{equation}
that is,
\begin{equation}\label{eq:mu23}
\mu^{\pm 1}=\frac{(p-1)\pm\sqrt{(p-1)^{2}-4}}{2}.
\end{equation}
The notation $\mu^{\pm1}$ reflects the fact that the product of these two roots equals one.
\end{proposition}
\begin{proof}
Write the characteristic polynomial of $\mathbb M$ as
\begin{equation}\label{eq:charpoly}
\chi(\mu)=\det\bigl(\mu\mathbb I-\mathbb M\bigr)=\mu^{3}-p\,\mu^{2}+q\,\mu-\det\mathbb M ,
\qquad q=\frac{1}{2}\bigl[(\tr\mathbb M)^{2}-\tr(\mathbb M^{2})\bigr],
\end{equation}
where $q$ is the sum of the principal $2\times2$ minors of $\mathbb M$. By Lemma~\ref{lem:det} we have $\det\mathbb M=1$, so that $\mu_1\mu_2\mu_3=1$ and hence
\[
q=\mu_1\mu_2+\mu_1\mu_3+\mu_2\mu_3=\frac{1}{\mu_3}+\frac{1}{\mu_2}+\frac{1}{\mu_1}=\tr\bigl(\mathbb M^{-1}\bigr).
\]
On the other hand, the reversibility relation \eqref{eq:rev} states that $\mathbb M=S\,\mathbb M^{-1}S$ with $S=\mathrm{diag}(1,-1,1)$, $S^{2}=\mathbb I$. Using the cyclic property of the trace,
\[
p=\tr\mathbb M=\tr\bigl(S\,\mathbb M^{-1}S\bigr)=\tr\bigl(\mathbb M^{-1}S\,S\bigr)=\tr\bigl(\mathbb M^{-1}\bigr)=q .
\]
Therefore
\begin{equation}\label{eq:charfactor}
\chi(\mu)=\mu^{3}-p\,\mu^{2}+p\,\mu-1=(\mu-1)\bigl(\mu^{2}-(p-1)\,\mu+1\bigr),
\end{equation}
as one checks by expanding the right-hand side. This proves \eqref{eq:mu1} and \eqref{eq:mu23eq}, while \eqref{eq:mu23} follows from the quadratic formula; the product of the two roots equals the constant term, whence $\mu_{2}\mu_{3}=1$. We may therefore write $\mu_2=\mu$ and $\mu_3=\mu^{-1}$.
\end{proof}

\begin{remark}[The multiplier $1$ and the conserved quantity]\label{rem:mult1}
Proposition~\ref{prop:factor} shows that $\mu_1=1$ is always a Floquet multiplier. The structural reason is the existence of the conserved quantity \eqref{eq:I}, which we write as
\begin{equation}\label{I scalar product}
\mathcal I=\mathbf w(\theta)^T\boldsymbol{\mathcal E}(\theta),\qquad
\mathbf w(\theta):=\begin{pmatrix}\nu^2\,\mathbf p(\theta)\\[2pt] -2\ee^2\end{pmatrix}=\begin{pmatrix}\nu^2(1-\ee\cos\theta)\\-\nu^2\ee\sin\theta\\ -2\ee^2\end{pmatrix}.
\end{equation}
The vector $\mathbf w$ is $2\pi$-periodic and never vanishes, since its last component equals $-2\ee^2\neq0$.\\
\noindent
Every solution is $\boldsymbol{\mathcal E}(\theta)=\F(\theta)\boldsymbol{\mathcal E}(0)$, and $\mathcal I$ is conserved. Therefore, for every choice of $\boldsymbol{\mathcal E}(0)\in\mathbb R^3$ we have
\begin{equation}\label{eq:leftinv}
\mathbf w(\theta)^T\boldsymbol{\mathcal E}(\theta)=\mathbf w(\theta)^T\left(\mathbb{F}(\theta)\boldsymbol{\mathcal E}(0)\right)=\mathcal{I}(\theta)=\mathcal{I}(0)=\mathbf w(0)^T\boldsymbol{\mathcal E}(0)  \  \Rightarrow\ \mathbf w(\theta)^T\,\F(\theta)=\mathbf w(0)^T\quad\forall\theta .
\end{equation}
Evaluating \eqref{eq:leftinv} at $\theta=2\pi$ and using $\mathbf w(2\pi)=\mathbf w(0)$ gives
\begin{equation}\label{eq:lefteig}
\mathbf w(0)^T\,\mathbb M=\mathbf w(0)^T ,
\end{equation}
so $\mathbf w(0)$ is a left eigenvector of the monodromy matrix for the eigenvalue $1$. Since a matrix and its transpose have the same spectrum, this recovers $\mu_1=1$ without using the reversibility of Lemma~\ref{lem:rev}.\\
\noindent
The converse holds as well. If $\mathbf{\bar{w}}^T\mathbb M=\mathbf{\bar{w}}^T$ for some $\mathbf{\bar{w}}\neq0$, set $\mathbf w(\theta)^T:=\mathbf{\bar{w}}^T\F(\theta)^{-1}$. Then $\mathbf w(\theta)^T\boldsymbol{\mathcal E}(\theta)=\mathbf{\bar{w}}^T\boldsymbol{\mathcal E}(0)$; since the right-hand side does not depend on $\theta$, $\mathbf w(\theta)^T\boldsymbol{\mathcal E}(\theta)$ is conserved, and $\mathbf w$ is $2\pi$-periodic, because $\F(\theta+2\pi)^{-1}=\mathbb M^{-1}\F(\theta)^{-1}$ and $\mathbf{\bar{w}}^T\mathbb M^{-1}=\mathbf{\bar{w}}^T$. A Floquet multiplier equal to $1$ and a $2\pi$-periodic linear first integral are thus two expressions of the same fact. In the problem at hand, this first integral is the linearised potential vorticity $\mathcal I$.
\end{remark}

\begin{corollary}\label{cor:discriminant}
Let $\ee\in(0,1)$ and let $\{1,\mu,\mu^{-1}\}$ be the Floquet multipliers of \eqref{eq:Asys}. Set
\begin{equation}\label{eq:Delta}
\Delta(\ee,\nu^2):=\tr\mathbb M-1=\mu+\mu^{-1}\in\mathbb R .
\end{equation}
Then the amplitudes $[\mathfrak{U}_1,\mathfrak{U}_3,\widetilde{\B}]$ (equivalently $(\A,\B)$) grow exponentially in $t$ if and only if $|\Delta|>2$, in which case the growth rate is
\begin{equation}\label{eq:sigma}
\sigma=\frac{\Omega}{2\pi}\,\arcosh\frac{|\Delta|}{2}.
\end{equation}
If $|\Delta|<2$ the amplitudes remain bounded, while if $|\Delta|=2$ there is no exponential growth, but the amplitudes may grow algebraically.
\end{corollary}
\begin{proof}
By Proposition~\ref{prop:factor}, $\tr\mathbb M=1+\mu+\mu^{-1}$, which is \eqref{eq:Delta}; the number $\Delta$ is real because $\mathbb M$ is a real matrix. The two multipliers $\mu^{\pm1}$ are the roots of $\lambda^2-\Delta\lambda+1=0$, that is
\begin{equation}\label{eq:quad}
\mu^{\pm 1}=\frac{\Delta\pm\sqrt{\Delta^2-4}}{2},
\end{equation}
and three cases arise.
\begin{itemize}
\item If $|\Delta|>2$, the roots are real and distinct. They have the sign of $\Delta$, because their product is positive, and, being reciprocal, exactly one of them lies outside the unit circle:
\[
|\mu_*|:=\max\{|\mu|,|\mu|^{-1}\}=\frac{|\Delta|+\sqrt{\Delta^2-4}}{2}>1 .
\]
\item If $|\Delta|<2$, the roots are complex conjugate with $|\mu|^2=\mu\cdot\mu^{-1}=1$; hence $\mu^{\pm1}=e^{\pm i\varphi}$ with $\Delta=2\cos\varphi$ and $\varphi\in(0,\pi)$.
\item If $|\Delta|=2$, there is a double root $\mu=\mu^{-1}=\Delta/2=\pm1$.
\end{itemize}

Assume now $|\Delta|>2$. By Remark~\ref{rem:time} the one-period map in physical time is similar to $\mathbb M^{-1}$, so we may work with the latter. It has the real eigenvalue $\mu_*$ with $|\mu_*|>1$ and, since $\mathbb M^{-1}-\mu_*\mathbb I$ is real and singular, a real eigenvector $\mathbf v\in\mathbb R^3\setminus\{0\}$. Choosing $\boldsymbol{\mathcal E}(0)=\F(ka)\mathbf v$ gives $\boldsymbol{\mathcal E}(t)=\F(\theta(t))\mathbf v$ and, by \eqref{eq:oneperiodmap},
\[
\boldsymbol{\mathcal E}(t+T)=\F(\theta(t))\,\mathbb M^{-1}\mathbf v=\mu_*\,\boldsymbol{\mathcal E}(t).
\]
Writing $t=nT+s$ with $n\in\mathbb N_0$ and $s\in[0,T)$, we obtain $\boldsymbol{\mathcal E}(t)=\mu_*^{\,n}\boldsymbol{\mathcal E}(s)$ and $|\boldsymbol{\mathcal E}(t)|=|\mu_*|^{\,n}|\boldsymbol{\mathcal E}(s)|$; note that $\mu_*<0$ whenever $\Delta<-2$, in which case the amplitude changes sign at every period. The function $s\mapsto|\boldsymbol{\mathcal E}(s)|=|\F(\theta(s))\mathbf v|$ is continuous and nonvanishing on the compact interval $[0,T]$, because $\F$ is invertible and $\mathbf v\neq0$, hence bounded between two positive constants $d_1\le d_2$. Setting
\[
\sigma:=\frac{\ln|\mu_*|}{T}>0,
\]
we have $|\mu_*|^{\,n}=e^{\sigma(t-s)}$ with $t-s=nT$, and since $s\in[0,T)$,
\[
d_1\,e^{-\sigma T}\,e^{\sigma t}\ \le\ |\boldsymbol{\mathcal E}(t)|\ \le\ d_2\,e^{\sigma t}\qquad(t\ge0),
\]
that is, $\boldsymbol{\mathcal E}$ grows exactly at the rate $\sigma$, in the sense that $t^{-1}\ln|\boldsymbol{\mathcal E}(t)|\to\sigma$.

This rate is maximal. Indeed $\mu$ is real with $\mu\neq\pm1$ when $|\Delta|>2$, so the three multipliers $1,\mu,\mu^{-1}$ are pairwise distinct and $\mathbb M^{-1}$ is diagonalisable, whence $\|\mathbb M^{-n}\|\le C|\mu_*|^{\,n}$. Writing $\theta(t)=ka-2\pi n-\Omega s$ for an arbitrary solution,
\[
\boldsymbol{\mathcal E}(t)=\F(ka-\Omega s)\,\mathbb M^{-n}\,\F(ka)^{-1}\boldsymbol{\mathcal E}(0),
\]
and the two outer factors are bounded, $s$ ranging over a compact interval; hence $|\boldsymbol{\mathcal E}(t)|\le C'e^{\sigma t}|\boldsymbol{\mathcal E}(0)|$ for every solution. The same argument proves the remaining statements: if $|\Delta|<2$ the three multipliers $1,e^{\pm i\varphi}$ are pairwise distinct and of modulus one, so $\|\mathbb M^{-n}\|$ is bounded uniformly in $n$ and every solution is bounded; if $|\Delta|=2$ all multipliers have modulus one, so $\|\mathbb M^{-n}\|$ grows at most polynomially and no solution grows exponentially.

Finally, $\arcosh x=\ln\bigl(x+\sqrt{x^{2}-1}\bigr)$ for $x\ge1$, so that
\[
\sigma=\frac{\Omega}{2\pi}\ln|\mu_*|
=\frac{\Omega}{2\pi}\ln\left(\frac{|\Delta|}{2}+\sqrt{\frac{\Delta^{2}}{4}-1}\right)
=\frac{\Omega}{2\pi}\arcosh\frac{|\Delta|}{2},
\]
which is \eqref{eq:sigma}.
\end{proof}
\noindent
Corollary~\ref{cor:discriminant} reduces the entire stability question to the sign and size of a single real function $\Delta(\ee,\nu^2)$. The conserved quantity of Proposition~\ref{prop:PV} localises, in addition, the unstable dynamics in phase space.

\begin{lemma}\label{lem:plane}
Let $\mathbf w(\theta)$ be as in \eqref{I scalar product}, and denote
\[
\mathbf w:=\mathbf w(0)=\left(\nu^{2}(1-\ee),\, 0,\, -2\ee^{2}\right)^T\neq\mathbf 0 .
\]
For $\ell\in\mathbb R$ and $\theta\in\mathbb R$ set $\Pi^\ell_\theta:=\{\boldsymbol{\mathcal E}\in\mathbb R^3:\ \mathbf w(\theta)^T\boldsymbol{\mathcal E}=\ell\}$ and $\Pi_\theta:=\Pi^0_\theta$. Thus $\Pi^\ell_\theta$ is the plane of the amplitudes whose potential vorticity \eqref{eq:I} equals $\ell$, and $\Pi_0$ is the plane of the initial amplitudes carrying no potential vorticity. Then:
\begin{enumerate}
\item[(i)] $\F(\theta)\,\Pi^\ell_0=\Pi^\ell_\theta$ for all $\theta,\ell\in\mathbb R$; since $\mathbf w$ is $2\pi$-periodic, $\mathbb M$ maps each plane $\Pi^\ell_0$ onto itself, and in particular $\mathbb M\,\Pi_0=\Pi_0$;
\item[(ii)] the restriction ${\mathsf M}_0:=\mathbb M|_{\Pi_0}$ has $\det{\mathsf M}_0=1$ and $\tr{\mathsf M}_0=\Delta$, and its eigenvalues are $\mu$ and $\mu^{-1}$;
\item[(iii)] if $\Delta\neq2$, the multiplier $1$ is simple and its eigenvector $\mathbf v_1$ satisfies $\mathbf w^{T}\mathbf v_1\neq0$, so that $\mathbf v_1$ is transverse to $\Pi_0$ and $\mathbb R^{3}=\Pi_0\oplus\operatorname{span}\{\mathbf v_1\}$. Moreover, $\F(\theta)\mathbf v_1$ is a $2\pi$-periodic solution; for every $\ell$ the plane $\Pi^\ell_0$ contains exactly one vector fixed by $\mathbb M$, namely $\boldsymbol{\mathcal E}^*_\ell=\ell\,\boldsymbol{\mathcal E}^*_1$ with $\boldsymbol{\mathcal E}^*_1:=\mathbf v_1/(\mathbf w^T\mathbf v_1)$; and every solution with $\mathcal I=\ell$ is the sum of the periodic solution through $\boldsymbol{\mathcal E}^*_\ell$ and of a solution lying in the planes $\Pi_\theta$;
\item[(iv)] if $|\Delta|>2$, the multipliers $\mu^{\pm1}$ are real and their eigenvectors lie in $\Pi_0$; if $|\Delta|<2$, the matrix ${\mathsf M}_0$ is conjugate to a rotation and every solution starting in $\Pi_0$ stays bounded.
\end{enumerate}
In other words, the growing and the decaying Floquet modes carry no potential-vorticity perturbation, whereas the neutral one does.
\end{lemma}
\begin{proof}
\begin{enumerate}
\item[(i)] Let $\boldsymbol{\mathcal E}(0)\in\Pi^\ell_0$. By Proposition~\ref{prop:PV}, $\mathbf w(\theta)^T\boldsymbol{\mathcal E}(\theta)=\mathbf w(0)^T\boldsymbol{\mathcal E}(0)=\ell$, so $\F(\theta)\boldsymbol{\mathcal E}(0)\in\Pi^\ell_\theta$ and $\F(\theta)\Pi^\ell_0\subseteq\Pi^\ell_\theta$. Conversely, given $\boldsymbol{\mathcal E}\in\Pi^\ell_\theta$, the solution with initial value $\F(\theta)^{-1}\boldsymbol{\mathcal E}$ takes the value $\boldsymbol{\mathcal E}$ at $\theta$, and the same conservation law gives $\mathbf w(0)^T\F(\theta)^{-1}\boldsymbol{\mathcal E}=\mathbf w(\theta)^T\boldsymbol{\mathcal E}=\ell$, i.e.\ $\F(\theta)^{-1}\boldsymbol{\mathcal E}\in\Pi^\ell_0$. Finally, since $\mathbf p$ is $2\pi$-periodic, $\mathbf w(2\pi)=\mathbf w(0)$, hence $\Pi^\ell_{2\pi}=\Pi^\ell_0$ and $\mathbb M\,\Pi^\ell_0=\F(2\pi)\,\Pi^\ell_0=\Pi^\ell_0$.
\item[(ii)] Choose a basis $\{\mathbf v'_1,\mathbf v'_2\}$ of $\Pi_0$ and a vector $\mathbf v'_3$ with $\mathbf w^T\mathbf v'_3=1$. If $\boldsymbol{\mathcal E}=a_1\mathbf v'_1+a_2\mathbf v'_2+a_3\mathbf v'_3$, then $\mathbf w^T\boldsymbol{\mathcal E}=a_3$: in these coordinates the third component of a vector is precisely its value of $\mathcal I$. By (i), $\mathbb M\mathbf v'_1$ and $\mathbb M\mathbf v'_2$ lie in $\Pi_0$, so their third components vanish; by \eqref{eq:lefteig}, $\mathbf w^T\mathbb M\mathbf v'_3=\mathbf w^T\mathbf v'_3=1$, so the third component of $\mathbb M\mathbf v'_3$ equals $1$. Hence the matrix of $\mathbb M$ in this basis is
\[
\begin{pmatrix}{\mathsf M}_0&\mathbf m\\ 0&1\end{pmatrix},\qquad \mathbf m\in\mathbb R^2,
\]
where ${\mathsf M}_0$ is the matrix of $\mathbb M|_{\Pi_0}$ in the basis $\{\mathbf v'_1,\mathbf v'_2\}$, and therefore $\det(z\mathbb I-\mathbb M)=(z-1)\det(z\mathbb I-{\mathsf M}_0)$. Comparing with \eqref{eq:charfactor}, namely $\det(z\mathbb I-\mathbb M)=(z-1)(z^{2}-\Delta z+1)$, gives $\det(z\mathbb I-{\mathsf M}_0)=z^{2}-\Delta z+1$, that is $\tr{\mathsf M}_0=\Delta$, $\det{\mathsf M}_0=1$, and the eigenvalues of ${\mathsf M}_0$ are $\mu^{\pm1}$.
\item[(iii)] Assume $\Delta\neq2$. The quadratic factor $z^2-\Delta z+1$ in \eqref{eq:charfactor} vanishes at $z=1$ if and only if $2-\Delta=0$. Hence, for $\Delta\neq2$, $z=1$ is a simple root of the characteristic polynomial of $\mathbb M$, and the vectors fixed by $\mathbb M$ form a line: $\{\boldsymbol{\mathcal E}:\mathbb M\boldsymbol{\mathcal E}=\boldsymbol{\mathcal E}\}=\operatorname{span}\{\mathbf v_1\}$, with $\mathbf v_1\neq\mathbf 0$.\\
By (ii), the eigenvalues of ${\mathsf M}_0$ are $\mu^{\pm1}$, and neither equals $1$ because $\Delta\neq2$. Hence no nonzero vector of $\Pi_0$ is fixed by $\mathbb M$: if $\mathbf u\in\Pi_0$ and $\mathbb M\mathbf u=\mathbf u$, then ${\mathsf M}_0\mathbf u=\mathbf u$ and so $\mathbf u=\mathbf 0$. Since $\mathbf v_1\neq\mathbf 0$ is fixed, $\mathbf v_1\notin\Pi_0$, i.e.\ $\mathbf w^T\mathbf v_1\neq0$. As $\Pi_0$ is a plane not containing $\mathbf v_1$, we get $\mathbb R^3=\Pi_0\oplus\operatorname{span}\{\mathbf v_1\}$; explicitly, every $\boldsymbol{\mathcal E}\in\mathbb R^3$ splits as
\[
\boldsymbol{\mathcal E}=\frac{\mathbf w^T\boldsymbol{\mathcal E}}{\mathbf w^T\mathbf v_1}\,\mathbf v_1+\Big(\boldsymbol{\mathcal E}-\frac{\mathbf w^T\boldsymbol{\mathcal E}}{\mathbf w^T\mathbf v_1}\,\mathbf v_1\Big),
\]
where the second term lies in $\Pi_0$ because $\mathbf w^T$ annihilates it.\\ By \eqref{eq:period}, $\F(\theta+2\pi)\mathbf v_1=\F(\theta)\,\mathbb M\mathbf v_1=\F(\theta)\mathbf v_1$ for all $\theta$, so the solution starting at $\mathbf v_1$ is $2\pi$-periodic.
\\ Since $1$ is a simple root of the characteristic polynomial of $\mathbb M$, a vector is fixed by $\mathbb M$ if and only if it equals $\tau\,\mathbf v_1$ for some $\tau\in\mathbb R$, and $\tau\,\mathbf v_1\in\Pi^\ell_0$ if and only if $\tau\,\mathbf w^T\mathbf v_1=\ell$. Since $\mathbf w^T\mathbf v_1\neq0$, this equation has exactly one solution $\tau=\ell/(\mathbf w^T\mathbf v_1)$. Hence $\Pi^\ell_0$ contains exactly one fixed vector of $\mathbb M$, namely
\[
\boldsymbol{\mathcal E}^*_\ell=\frac{\ell}{\mathbf w^T\mathbf v_1}\,\mathbf v_1=\ell\,\boldsymbol{\mathcal E}^*_1,\qquad \boldsymbol{\mathcal E}^*_1:=\frac{\mathbf v_1}{\mathbf w^T\mathbf v_1};
\]
geometrically, it is the intersection of the line $\operatorname{span}\{\mathbf v_1\}$ with the plane $\Pi^\ell_0$, which is a single point because the line is transverse to the plane.\\ Finally, let $\boldsymbol{\mathcal E}(\theta)$ be a solution with $\mathcal I=\ell$, i.e.\ $\boldsymbol{\mathcal E}(0)\in\Pi^\ell_0$. Then $\mathbf w^T\big(\boldsymbol{\mathcal E}(0)-\boldsymbol{\mathcal E}^*_\ell\big)=\ell-\ell=0$, so $\boldsymbol{\mathcal E}(0)-\boldsymbol{\mathcal E}^*_\ell\in\Pi_0$, and by linearity
\[
\boldsymbol{\mathcal E}(\theta)=\F(\theta)\boldsymbol{\mathcal E}^*_\ell+\F(\theta)\big(\boldsymbol{\mathcal E}(0)-\boldsymbol{\mathcal E}^*_\ell\big).
\]
The first term is $2\pi$-periodic, being a multiple of $\F(\theta)\mathbf v_1$, and carries the whole potential vorticity $\ell$; the second term lies in $\Pi_\theta$ for every $\theta$ by (i) with $\ell=0$, and carries no potential vorticity.
\item[(iv)] If $|\Delta|>2$, the roots $\mu^{\pm1}$ are real, distinct and different from $1$. For such a root, $\mathbb M-\mu\mathbb I$ is a real singular matrix, so it has a real eigenvector $\mathbf v$, and from $\mathbf w^{T}\mathbf v=\mathbf w^{T}\mathbb M\mathbf v=\mu\,\mathbf w^{T}\mathbf v$ and $\mu\neq1$ we get $\mathbf w^{T}\mathbf v=0$, i.e.\ $\mathbf v\in\Pi_0$. If $|\Delta|<2$, then ${\mathsf M}_0$ is a real $2\times2$ matrix with $\det{\mathsf M}_0=1$ and $|\tr{\mathsf M}_0|<2$; its eigenvalues are $e^{\pm i\varphi}$ with $2\cos\varphi=\Delta$ and $\varphi\in(0,\pi)$, so it is conjugate to the rotation by the angle $\varphi$ and its powers are bounded. Writing $\theta=2\pi n+s$ with $n\in\mathbb Z$ and $s\in[0,2\pi)$, any solution starting at $\boldsymbol{\mathcal E}(0)\in\Pi_0$ satisfies $\boldsymbol{\mathcal E}(\theta)=\F(s)\,\mathbb M^n\boldsymbol{\mathcal E}(0)=\F(s)\,{\mathsf M}_0^n\boldsymbol{\mathcal E}(0)$, which is bounded since $\F$ is bounded on $[0,2\pi]$.
\end{enumerate}
\end{proof}
\noindent
The first two components of the normal $\mathbf w(\theta)$ are $\nu^2\mathbf p(\theta)=\nu^2\big[(1,0)-\ee(\cos\theta,\sin\theta)\big]$, which traverse once per orbit the circle of centre $(\nu^2,0)$ and radius $|\nu^2|\ee$, while the third component $-2\ee^2$ is fixed. The plane $\Pi_\theta$ therefore wobbles periodically, returning to $\Pi_0$ after each orbital period; for $\nu=0$ the normal is constant and the planes reduce to $\widetilde{\B}=\mathrm{const}$. Thus the three-dimensional amplitude dynamics splits into the dynamics on $\Pi_\theta$, carrying the multipliers $\mu^{\pm1}$, and a neutral direction spanned by $\boldsymbol{\mathcal E}^*_1$, carrying the multiplier $1$ and the potential vorticity.

\subsubsection{Reduction to a Hill equation}
Since, as stated above, the instability of the system is connected to $|\Delta|$, it is useful to have an even more explicit (scalar, second-order) form of the problem, both to compute $\Delta$ numerically and to make contact with the classical Mathieu/Hill instability literature. Decompose $\mathbf V=(\mathfrak{U}_1,\mathfrak{U}_3)^T$ along and across the isentropes using $\mathbf p$ and $\mathbf q$:
\[
A:=\mathbf p\cdot\mathbf V,\qquad \beta:=\frac{\mathbf q\cdot\mathbf V}{h},\qquad\text{so that}\qquad \mathbf V=\frac{A\,\mathbf p+h\beta\,\mathbf q}{h}.
\]
Besides $\mathbf p'=-\Msf\mathbf p$, we use $\mathbf q'=\Msf^T\mathbf q$. This identity expresses the fact that $\nabla b=\mathbf q/\D$ is transported by the flow, and it can be checked directly: $\Msf^T\mathbf q=\ee(\cos\theta,\sin\theta)^T=\mathbf q'$. Then \eqref{eq:pVprime} gives $A'=-\frac{2\ee^2}{\D}\,\mathbf q\cdot\mathbf V$, while \eqref{eq:compact} gives $(\mathbf q\cdot\mathbf V)'=(\mathbf q'+\Msf^T\mathbf q)\cdot\mathbf V+h\widetilde{\B}=2\,\mathbf q'\cdot\mathbf V+h\widetilde{\B}$. Using $\mathbf q'\cdot\mathbf p=\ee(\cos\theta-\ee)$, $\mathbf q'\cdot\mathbf q=\ee\sin\theta$ and $h'=2\ee\sin\theta$, we obtain
\begin{equation}\label{eq:Abeta}
A'=-\frac{2\ee^2h}{\D}\,\beta,\qquad
\beta'=\frac{2\ee(\cos\theta-\ee)}{h^2}\,A+\widetilde{\B},\qquad
\widetilde{\B}'=-\frac{\nu^2h}{\D}\,\beta .
\end{equation}
Since $\ee>0$, we may eliminate $\beta=-\frac{\D}{2\ee^2}\frac{A'}{h}$ and $\widetilde{\B}=(\nu^2A-\mathcal I)/(2\ee^2)$. This yields
\begin{equation}\label{eq:Hill}
\left(\frac{A'}{h}\right)'+Q(\theta)\,A=\frac{\mathcal I}{\D},\qquad
Q(\theta)=\frac{1}{\D}\left[\nu^2-\frac{4\ee^3(\ee-\cos\theta)}{h(\theta)^2}\right].
\end{equation}
On the plane $\Pi_0$ of Lemma~\ref{lem:plane}, where the perturbation carries no potential vorticity, the constant $\mathcal I$ vanishes and \eqref{eq:Hill} reduces to the homogeneous Hill equation $(A'/h)'+QA=0$, written in Sturm--Liouville form; in the variable $\tau=\int_0^{\theta}h(s)\,{\rm d}s$, it takes the standard form ${\rm d}^{2}A/{\rm d}\tau^{2}+(Q/h)A=0$. The following proposition makes the correspondence precise, by analysing the Hill equation on the invariant plane.

\begin{proposition}\label{prop:hill}
For each $\theta$, the map $\Pi_\theta\ni(\mathbf V, \widetilde{\B})\longmapsto(A,B)$ with
\[
A=\mathbf p\cdot\mathbf V,\qquad B=-\frac{2\ee^2}{\D\,h}\,\mathbf q\cdot\mathbf V
\]
is a linear isomorphism from $\Pi_\theta$ onto $\mathbb R^2$. Along solutions lying in the planes $\Pi_\theta$, the pair $(A,B)$ satisfies
\begin{equation}\label{eq:Hillsys}
A'=h\,B,\qquad B'=-Q\,A,\qquad\text{with}\qquad Q= \frac1\D\Big[\nu^2-\frac{4\ee^3(\ee-\cos\theta)}{h^2}\Big]
\end{equation}
which is the homogeneous Hill equation $(A'/h)'+QA=0$; its Floquet discriminant equals $\Delta$.
\end{proposition}
\begin{proof}
On $\Pi_\theta$ the condition $\mathcal I=0$ gives $\widetilde{\B}=\frac{\nu^2}{2\ee^2}\,\mathbf p\cdot\mathbf V$, so that $\mathbf V\mapsto(\mathbf V,\frac{\nu^2}{2\ee^2}\mathbf p\cdot\mathbf V)$ is an isomorphism from $\mathbb R^2$ onto $\Pi_\theta$ and $\mathbf V$ is a coordinate on the plane. The map $\mathbf V\mapsto(A,B)$ has matrix
\[
\Lsf(\theta)=\begin{pmatrix}\mathbf p(\theta)^T\\[2pt] -\dfrac{2\ee^2}{\D\,h(\theta)}\,\mathbf q(\theta)^T\end{pmatrix},\qquad
\det \Lsf(\theta)=-\frac{2\ee^2}{\D\,h}\,(p_1q_2-p_2q_1)=-\frac{2\ee^2}{\D}\neq0,
\]
since $\mathbf q=J\mathbf p$ gives $p_1q_2-p_2q_1=|\mathbf p|^2=h$. Along every solution, $(\mathbf p\cdot\mathbf V)'=-\frac{2\ee^2}{\D}\,\mathbf q\cdot\mathbf V$ by \eqref{eq:pVprime}, i.e.\ $A'=hB$. Moreover, by \eqref{eq:Abeta}, $\beta=\mathbf q\cdot\mathbf V/h$ satisfies $\beta'=\frac{2\ee(\cos\theta-\ee)}{h^2}A+\widetilde{\B}$, and on $\Pi_\theta$ we have $\widetilde{\B}=\frac{\nu^2}{2\ee^2}A$; since $B=-\frac{2\ee^2}{\D}\beta$,
\[
B'=-\frac{2\ee^2}{\D}\Big[\frac{2\ee(\cos\theta-\ee)}{h^2}+\frac{\nu^2}{2\ee^2}\Big]A=-\frac1\D\Big[\nu^2-\frac{4\ee^3(\ee-\cos\theta)}{h^2}\Big]A=-QA .
\]
For the discriminant, let $\mathsf{H}(\theta)$ be the fundamental matrix of \eqref{eq:Hillsys} and let $\widehat\F(\theta)$ be the $2\times2$ matrix of $\F(\theta)\colon\Pi_0\to\Pi_\theta$ in the coordinates $\mathbf V$. Every solution satisfies $(A,B)(\theta)=\Lsf(\theta)\widehat\F(\theta)\Lsf(0)^{-1}(A,B)(0)$, hence $\mathsf{H}(\theta)=\Lsf(\theta)\widehat\F(\theta)\Lsf(0)^{-1}$. Since $\mathsf{L}$ is $2\pi$-periodic, $\mathsf{H}(2\pi)=\Lsf(0)\widehat\F(2\pi)\Lsf(0)^{-1}$ is similar to $\widehat\F(2\pi)$, which is the matrix of ${\mathsf M}_0=\mathbb M|_{\Pi_0}$. Therefore $\tr \mathsf{H}(2\pi)=\tr{\mathsf M}_0=\Delta$ by Lemma~\ref{lem:plane}(ii).
\end{proof}
\noindent
This last Proposition therefore shows that the instability analysis for the system describing upward propagating mountain waves of \cite{ConstantinMountain2023} reduces to the analysis of a Hill equation and its discriminant.\\
The coefficient $Q$ acts as a stiffness that varies periodically in $\theta$, and its mean over one period can be computed in closed form. Since $\partial_\ee(1/h)=-2(\ee-\cos\theta)/h^2$, differentiating the elementary integral $\int_0^{2\pi}h^{-1}{\rm d}\theta=2\pi/\D$ with respect to $\ee$ gives
\[
\int_0^{2\pi}\frac{\ee-\cos\theta}{h(\theta)^{2}}\,{\rm d}\theta
=-\frac12\frac{\rm d}{{\rm d}\ee}\int_0^{2\pi}\frac{{\rm d}\theta}{h(\theta)}
=-\frac12\frac{\rm d}{{\rm d}\ee}\frac{2\pi}{\D}
=-\frac{2\pi\ee}{\D^{2}},
\]
whence
\begin{equation}\label{eq:Qmean}
\overline{Q}:=\frac{1}{2\pi}\int_0^{2\pi}Q\,{\rm d}\theta=\frac{1}{\D}\left(\nu^{2}+\frac{4\ee^{4}}{\D^{2}}\right)
\end{equation}
Hence, for $\nu^2\ge0$, the stiffness is restoring on average.

The sign of $\Delta$ distinguishes two instability mechanisms. In a static instability the stiffness is never restoring: $Q\le0$ on $[0,2\pi]$, with $Q\not\equiv0$. In this case $\Delta>2$, so that the multipliers are positive. Indeed, let $(A_1,B_1)$ and $(A_2,B_2)$ be the solutions of \eqref{eq:Hillsys} with $(A_1,B_1)(0)=(1,0)$ and $(A_2,B_2)(0)=(0,1)$, so that $\Delta=A_1(2\pi)+B_2(2\pi)$ by Proposition~\ref{prop:hill}. As long as $A_1>0$, we have $B_1'=-QA_1\ge0$, hence $B_1\ge0$ and $A_1'=hB_1\ge0$; thus $A_1\ge1$ on $[0,2\pi]$. Similarly, as long as $B_2>0$, we have $A_2'=hB_2>0$, hence $A_2>0$ and $B_2'=-QA_2\ge0$; thus $B_2\ge1$ on $[0,2\pi]$ and $A_2>0$ on $(0,2\pi]$. Since $Q<0$ on some open subinterval of $(0,2\pi)$, it follows that $B_2(2\pi)=1-\int_0^{2\pi}QA_2\,{\rm d}\theta>1$, and therefore $\Delta>2$. For the present coefficient, $Q=\D^{-1}\big[\nu^2-4\ee^3 f(\cos\theta)\big]$ with $f(c):=(\ee-c)/(1-2\ee c+\ee^2)^2$, and $f(c)\ge f(1)=-(1-\ee)^{-3}$ for all $c\in[-1,1]$: this is obvious for $c\le\ee$, where $f\ge0$, and follows for $c>\ee$ from $0<c-\ee\le1-\ee$ and $h=(1-\ee)^2+2\ee(1-c)\ge(1-\ee)^2$. Therefore $Q\le0$ on $[0,2\pi]$ if and only if
\[
\nu^2\le-\frac{4\ee^3}{(1-\ee)^3},
\]
and then $Q\not\equiv0$; this curve is shown in Figure~\ref{fig:stability}.\\
In a {parametric} instability, instead, the stiffness is restoring on average and the growth is produced by its periodic modulation. When, in addition, the multipliers are negative, i.e.\ $\Delta<-2$, as in the principal resonance of the Mathieu equation \cite{MagnusWinkler,McLachlan}, the growing solution reverses its sign at every period.

The instability found below is of the second kind. For $\nu^{2}\ge0$ the layer is statically stable, and so is the flat limit: by Theorem~\ref{prop:strat}(iii), $\Delta\to2\cos(2\pi\nu)\in(-2,2)$ as $\ee\to0$ whenever $2\nu\notin\mathbb Z$. The instability is therefore produced by the wave, not by convection. Moreover, it has $\Delta<-2$: the growing Floquet solution satisfies $A(\theta+2\pi)=\mu A(\theta)$ with $\mu<0$, so that it reverses its sign at every period and oscillates at half the orbital frequency $\Omega/2$, which is the signature of a subharmonic parametric resonance. This is consistent with the positive mean \eqref{eq:Qmean}, although the sign of the mean stiffness does not by itself decide stability: a positive mean is compatible with instability, as for the Mathieu equation \cite{MagnusWinkler,McLachlan}, and a negative mean with stability, as for Kapitza's inverted pendulum \cite{Kapitza1951,LeviWeckesser1995}.

Both signs of $\Delta$ occur in the problem: $\Delta>2$ for the convective instability ($\nu^2<0$) and, numerically, in the tongue issuing from $\nu=1$, whereas $\Delta<-2$ for the wave-induced instability (\S\ref{sec_neutral} and \S\ref{sec_strat}). Since $\Delta$ is continuous, it maps each connected component of $\{|\Delta|>2\}$ onto a connected subset of $(-\infty,-2)\cup(2,\infty)$, so that its sign is constant on each component. In particular, the component containing the segment $\{\ee>1/3,\ \nu=0\}$ satisfies $\Delta<-2$ throughout. Numerically, this component reaches the resonance tongue issuing from $(\ee,\nu)=(0,1/2)$, where the coefficients are only weakly modulated, at order $\ee$, and the classical Mathieu picture applies.

\subsubsection{Neutral stratification: an explicit solution}\label{sec_neutral}
When $\nu=0$ the amplitude $\widetilde{\B}$ is constant, the monodromy matrix is block triangular, $\mathbb M=\left(\begin{smallmatrix}X(2\pi)&*\\0&1\end{smallmatrix}\right)$ with $X$ the fundamental matrix of $\mathbf V'=\Msf\mathbf V$, and therefore $\Delta=\tr X(2\pi)$: the relevant dynamics is $\mathbf V'=\Msf\mathbf V$. Writing
\[
R_\alpha=\begin{pmatrix}
    \cos\alpha & -\sin\alpha\\
    \sin\alpha &\cos\alpha
\end{pmatrix}
\]
for the rotation by an angle $\alpha$, one checks that
\[
\Msf=\frac{\ee}{\D}\,R_{\alpha}\begin{pmatrix}1&0\\0&-1\end{pmatrix}R_{\alpha}^T+\frac{\ee^2}{\D}\,J,\qquad \alpha(\theta)=\frac{\theta}{2}+\frac{\pi}{4},
\]
that is, the principal axes of strain rotate at exactly half the orbital angular velocity. In the co-rotating frame $\mathbf V=R_{\alpha(\theta)}\mathbf W$ the system becomes autonomous:
\begin{equation}\label{eq:W}
\mathbf W'=\mathsf K\,\mathbf W,\qquad
\mathsf K=\begin{pmatrix}\eta&-\varpi\\ \varpi&-\eta\end{pmatrix},
\end{equation}
where
\begin{equation}\label{eq:gsome}
\eta:=\frac{\ee}{\D},\qquad\text{and}\qquad
\varpi:=\frac{\ee^{2}}{\D}-\frac12=\frac{3\ee^{2}-1}{2\D},
\end{equation}
the latter being the vorticity of the basic flow corrected by the rotation $\frac{1}{2}$ of the frame itself. The eigenvalues of $\mathsf K$ are $\pm\lambda$ with
\begin{equation}\label{eq:lambda}
\lambda^2=\eta^2-\varpi^2=\frac{9\ee^2-1}{4(1-\ee^2)},
\end{equation}
and the fundamental matrix of the system \eqref{eq:W} is $e^{\theta\mathsf K}$. From 
\begin{equation}
    \mathbf V(\theta)=R_{\alpha(\theta)}\mathbf W(\theta)=R_{\alpha(\theta)}e^{\theta\mathsf K}\mathbf W(0) =R_{\alpha(\theta)}e^{\theta\mathsf K}R_{\pi/4}^T \mathbf V(0)
\end{equation}
we note that the fundamental matrix of $\mathbf V'=\Msf\mathbf V$ is $X(\theta)=R_{\theta/2+\pi/4}\,e^{\theta\mathsf K}R_{\pi/4}^T$, so that the monodromy matrix $X(2\pi)=-R_{\pi/4}e^{2\pi\mathsf K}R_{\pi/4}^T$; its trace equals $-\tr e^{2\pi\mathsf K}$ by invariance of the trace under similarity. Finally, $\tr\mathsf K=0$ and $\det\mathsf K=-\lambda^2$, so the Cayley--Hamilton theorem gives $\mathsf K^2=\lambda^2\,\mathbb I$ and hence
\[
e^{\theta\mathsf K}=\cosh(\lambda\theta)\,\mathbb I+\frac{\sinh(\lambda\theta)}{\lambda}\,\mathsf K.
\]
Taking the trace at $\theta=2\pi$ we obtain
\begin{equation}\label{eq:Delta0}
\Delta(\ee,0)=-2\cosh(2\pi\lambda).
\end{equation}
The nature of the eigenvalues \eqref{eq:lambda} determines the behaviour of $\mathbf{W}$ (hence, of $\mathbf{V}$). If $\ee>1/3$, \eqref{eq:lambda} has two real solution---one positive and one negative---giving an exponential growth of the perturbation, leading to the instability of the problem. This is summarised in the following Theorem.

\begin{figure}
            \centering
            \includegraphics[width=\linewidth]{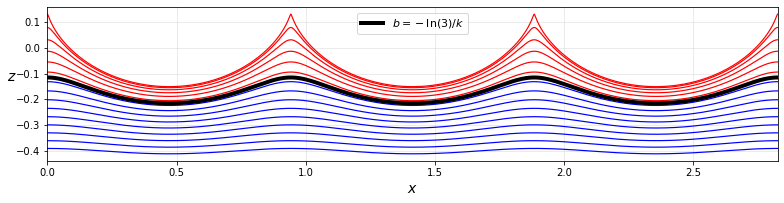}
            \caption{Trochoidal particle path of \eqref{solution} at fixed time $t=0$,  with $a \in [0,6\pi/k]$, $s=0$ and $b\in(-0.4,\,-0.01)$, with  $k=20/3$. In blue are depicted waves with $b<-\ln(3)/k$, and in black that for $b=-\ln(3)/k$. In red are depicted the unstable ones, for which $b>-\ln(3)/k$.}
            \label{fig:instab_gerstner}
        \end{figure}

\begin{theorem}\label{thm:neutral}
Let $\nu=0$ and $\ee=e^{kb}>1/3$, so that the eigenvalues of $\mathsf K$ are 
\begin{equation}
    \pm\lambda\in \mathbb{R} \qquad\text{with}\qquad\lambda=\frac{1}{2} \sqrt{\frac{9\ee^2-1}{1-\ee^2}}>0.
\end{equation}
We have the following
\begin{enumerate}
    \item[(i)] the short-wavelength perturbation at level $b$ grows exponentially, with growth rate
\begin{equation}\label{eq:sigma0}
\sigma=\Omega\,\lambda=\frac{\Omega}{2}\sqrt{\frac{9\ee^2-1}{1-\ee^2}},\qquad \Omega=\sqrt{gk};\end{equation}
\item[(ii)] $\Delta<-2$, namely the growing mode is subharmonic: it changes sign every orbital period, oscillating at $\Omega/2$;
\item[(iii)] the choice $\mathfrak{U}_2\equiv0$, $\B\equiv0$,
\begin{equation}\label{eq:explicit}
\begin{pmatrix}\mathfrak{U}_1\\\mathfrak{U}_3\end{pmatrix}(t)
=e^{-\lambda\,\theta(t)}\,R_{\theta(t)/2+\pi/4}\begin{pmatrix}\varpi\\\eta+\lambda\end{pmatrix},
\qquad \theta(t)=ka-\Omega t,
\end{equation}
is an exact solution of the WKB amplitude system \eqref{leading order} (restricted to $\boldsymbol\xi_0=(0,1,0)$) along the particle path with label $b=k^{-1}\ln\ee$, and $|(\mathfrak{U}_1,\mathfrak{U}_3)(t)|=|(\varpi,\eta+\lambda)|\,e^{\lambda(\Omega t-ka)}$ grows without bound as $t\to\infty$. 
\end{enumerate}
\end{theorem}
\begin{proof}
\begin{enumerate}
    \item[(i)] follows from \eqref{eq:Delta0} and Corollary~\ref{cor:discriminant}: for $\ee>1/3$, $\lambda>0$ is real and $|\Delta|=2\cosh(2\pi\lambda)>2$, giving \eqref{eq:sigma0} by \eqref{eq:sigma}.
\item[(ii)] the multipliers $-e^{\pm2\pi\lambda}$ are negative whenever $\ee>1/3$, giving the subharmonic character as in Corollary~\ref{cor:discriminant}.
\item[(iii)] An easy direct computation shows that the vector $\mathbf{z}=(\varpi,\ \eta+\lambda)^T$ satisfies $\mathsf K\mathbf{z}=-\lambda\mathbf{z}$.\\
By construction $\mathbf W(\theta)=e^{-\lambda\theta}(\varpi,\eta+\lambda)^T$ solves $\mathbf W'=\mathsf K\mathbf W$, being a multiple of the eigenvector for the eigenvalue $-\lambda$; then $\mathbf V(\theta)=R_{\alpha(\theta)}\mathbf W(\theta)$ solves $\mathbf V'=\Msf\mathbf V$ by the derivation of \eqref{eq:W}. Since $\theta$ decreases linearly with $t$ at rate $\Omega$, the factor $e^{-\lambda\theta(t)}$ grows like $e^{\lambda\Omega t}$, and $R_\alpha$, being a rotation, does not affect the modulus.
\end{enumerate}
\end{proof}
We stress that the sign of $\Delta$, and not only its modulus, carries physical information. The instability found here has $\Delta<-2$, i.e.\ negative Floquet multipliers: the growing perturbation reverses its sign at every orbit and oscillates at half the orbital frequency. The minus sign originates from the half turn performed by the principal axes of strain during one orbit, $R_{\alpha(2\pi)}=-R_{\alpha(0)}$, and, by continuity, it persists for $\nu\neq0$ (see \S\ref{sec_strat}).
\begin{remark}
    If $\ee=1/3$ the growth is algebraic (linear in $t$), whereas for $\ee<1/3$ every solution is bounded.\\
    Indeed at $\ee=1/3$ one has $\lambda=0$ and $\mathsf K=\frac{3}{8}\left(\begin{smallmatrix}1&1\\-1&-1\end{smallmatrix}\right)$ satisfies $\mathsf K^2=0$, so $e^{\theta\mathsf K}=\mathbb I+\theta\mathsf K$ grows linearly, and so does $X(\theta)$. \\For $\ee<1/3$, $\lambda=i|\lambda|$ with $0<|\lambda|<1/2$ is purely imaginary and $|\Delta|=2|\cos(2\pi|\lambda|)|<2$, so all multipliers lie on the unit circle and every solution is bounded.     
\end{remark}
 Since the particle kinematics \eqref{solution} coincide with those of Gerstner's wave, Theorem~\ref{thm:neutral} recovers the threshold $1/3$ found in \cite{Leblanc2004}.\\
 \\ \noindent
The neutral case $\nu\equiv0$ is not merely a limiting case: it is realised within the family of \cite{ConstantinMountain2023} by an explicit basic state. Indeed, take the isentropic relation \eqref{isentropic} with the same constant $K$ throughout the layer, $P=K\rho^{\gamma}$. Then $g_3=\ln\bigl(\rho P^{-1/\gamma}\bigr)=-\gamma^{-1}\ln K$ is constant, so that $\nu\equiv0$ by \eqref{eq:nu}. Combining $P_b=\gamma K\rho^{\gamma-1}\rho_b$ with $P_b=-g\rho\,(1-e^{2kb})$ gives
\[
\frac{\gamma K}{\gamma-1}\,\frac{{\rm d}}{{\rm d}b}\,\rho^{\gamma-1}=-g\,\bigl(1-e^{2kb}\bigr),
\]
and integrating from $b_1$ to $b$ we obtain
\begin{equation}\label{eq:isorho}
\rho(b)^{\gamma-1}=\rho(b_1)^{\gamma-1}-\frac{\kappa\, g}{K}\left[(b-b_1)-\frac{e^{2kb}-e^{2kb_1}}{2k}\right],\qquad \kappa:=\frac{\gamma-1}{\gamma}.
\end{equation}
The derivative of the bracket with respect to $b$ is $1-e^{2kb}>0$, so the density decreases with $b$, as required in \cite{ConstantinMountain2023}. It remains positive throughout the layer provided
\[
\rho(b_1)^{\gamma-1}>\frac{\kappa\,g}{K}\left[(b_0-b_1)-\frac{e^{2kb_0}-e^{2kb_1}}{2k}\right],
\]
that is, provided the layer is not too thick. Physically, since $T=P/\rho=K\rho^{\gamma-1}$, relation \eqref{eq:isorho} states that the temperature decreases as $T_b=-\kappa g\,(1-e^{2kb})$. Because $\nu\equiv0$ at every level, Theorem~\ref{thm:neutral} applies throughout the layer. The steepness $e^{kb}$ increases with $b$, so the layer $b_1<b<b_0$ contains levels with $e^{kb}>1/3$ if and only if $e^{kb_0}>1/3$. Hence this exact solution is linearly unstable whenever (see Fig.\ref{fig:instab_gerstner})
\begin{equation}\label{eq:b0isentropic}
b_0>-\frac{\ln3}{k}.
\end{equation}

\subsubsection{Stratified layers}\label{sec_strat}
\begin{figure}
\centering
\includegraphics[width=.85\linewidth]{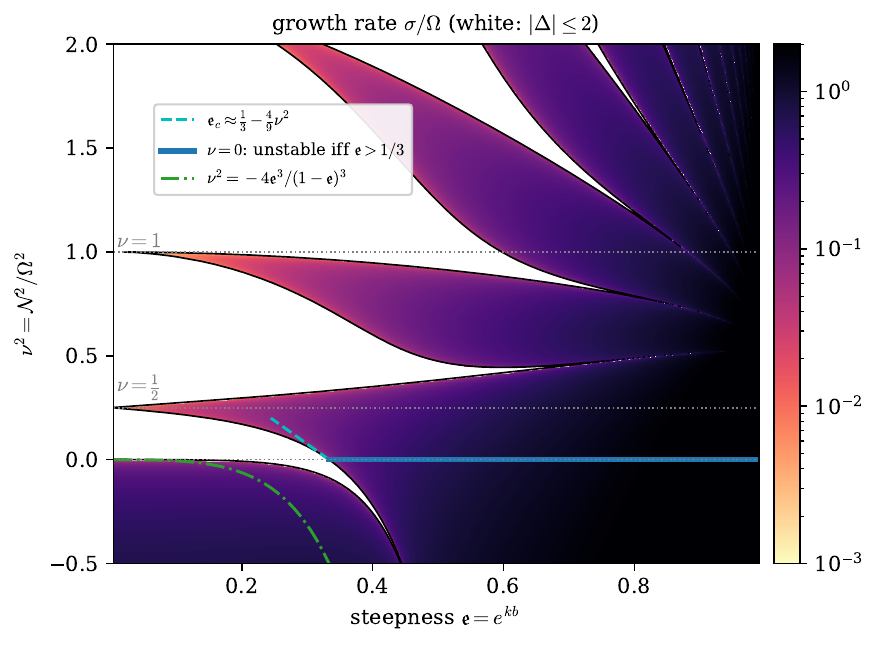}
\caption{Growth rate $\sigma/\Omega$ of short-wavelength perturbations in the $(\ee,\nu^2)$ plane (the coloured part being the unstable region where $|\Delta|>2$; white regions are not unstable ($|\Delta|\le2$). The thick blue segment is the exact result of Theorem~\ref{thm:neutral}, the dashed blue line the asymptotic neutral curve \eqref{eq:ec}. For $\nu^2<0$, the instability occur already for small steepness $\ee$, as stated in Theorem \ref{prop:strat}. The dashed green line is the region below which $Q\leq0$.}
\label{fig:stability}
\end{figure}

\begin{figure}
    \centering
    \includegraphics[width=.6\linewidth]{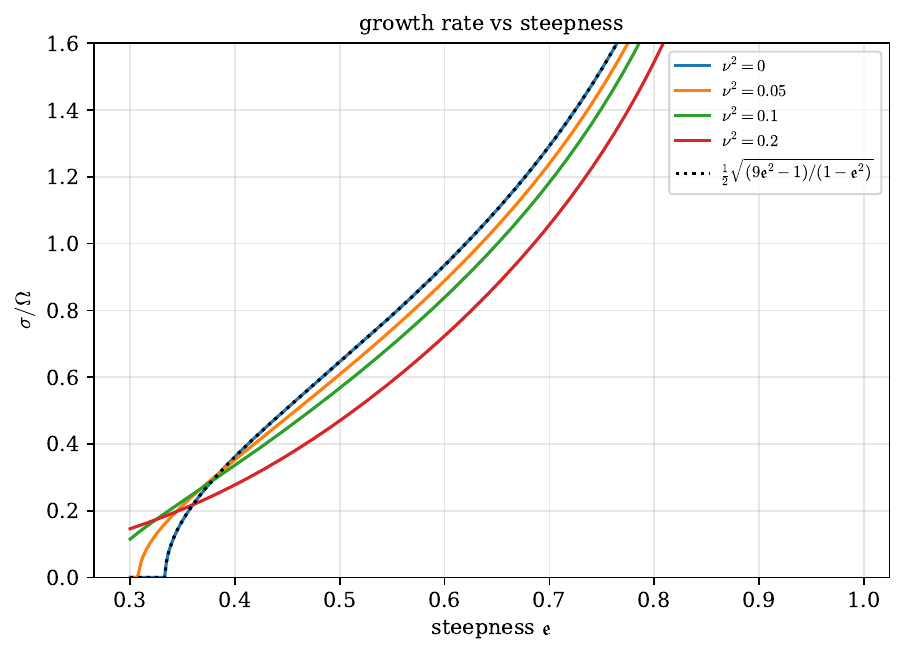}
    \caption{Growth rate as a function of the steepness for several values of $\nu^2$; the dotted line is \eqref{eq:sigma0}}
    \label{fig:rate}
\end{figure}

We now consider the case \(\nu\neq0\). Building upon the analysis for \(\nu=0\), we obtain the following characterization of the unstable set. The analytical description provided by the theorem is complemented, later in this subsection, by a numerical study (see Fig. \ref{fig:stability} and Fig. \ref{fig:rate}), which exhibits the corresponding instability structure in the parameter plane.

\begin{theorem}\label{prop:strat}
$\Delta(\ee,\nu^2)$, defined by \eqref{eq:Delta}, is real-analytic on $(0,1)\times\mathbb R$. Moreover:
\begin{enumerate}
\item[(i)] the unstable set $\mathcal S:=\{|\Delta|>2\}$ is open and contains the segment $\{(\ee,0):\ee>1/3\}$; hence for every $\ee\in(1/3,1)$ there is $\nu_*(\ee)>0$, possibly small, such that the level is unstable for all $|\nu^2|<\nu_*(\ee)^2$;
\item[(ii)] there are a neighbourhood of $(\ee,\nu^2)=(1/3,0)$ and a real-analytic function $\ee_c$ with $\ee_c(0)=1/3$ such that, in that neighbourhood, the level is unstable if and only if $\ee>\ee_c(\nu^2)$, and
\begin{equation}\label{eq:ec}
\ee_c(\nu^2)=\frac{1}{3}-\frac{4}{9}\,\nu^2+\mathcal{O}(\nu^4);
\end{equation}
\item[(iii)] at $\ee=0$ one has $\Delta=2\cos(2\pi\nu)$ for $\nu^2\ge0$ and $\Delta=2\cosh(2\pi|\nu|)$ for $\nu^2<0$; hence statically unstable layers (i.e. $\nu^2<0$) are unstable already for arbitrarily small steepness, while for $\nu^2>0$ the closure of $\mathcal S$ meets the axis $\ee=0$ only at the resonant values $\nu=n/2$, $n\in\mathbb N$.
\end{enumerate}
\end{theorem}

\begin{proof}
\begin{enumerate}
    \item [(i)] Analyticity of $\Delta$ follows from the analytic dependence of $\mathbb A(\theta;\ee,\nu^2)$ on the parameters and from the standard theory of linear ODEs with parameters (the fundamental matrix, hence its trace at $\theta=2\pi$, depends analytically on the coefficients); note that $\mathbb A$ is regular at $\ee=0$ as well. Since $\{|x|>2\}$ is open in $\mathbb R$ and $\Delta$ is continuous, $\mathcal S=\Delta^{-1}(\{|x|>2\})$ is open; it contains $\{\ee>1/3,\ \nu=0\}$ by Theorem~\ref{thm:neutral}. Openness gives, around each such point, a small disc contained in $\mathcal S$, in particular a vertical segment in $\nu^2$.
\item[(ii)] The neutral curve (i.e. the boundary between stable and unstable levels) near $(1/3,0)$ is the set where $\Delta=-2$. Since $\Delta(1/3,0)=-2$ by \eqref{eq:Delta0}, the implicit function theorem applied to $\Delta(\ee,\nu^2)+2=0$ describes this set as a graph $\ee=\ee_c(\nu^2)$, provided $\partial_\ee\Delta(1/3,0)\neq0$, and gives $\ee_c'(0)=-\partial_{\nu^2}\Delta/\partial_\ee\Delta$ at $(1/3,0)$. We compute the two derivatives in closed form.
\begin{itemize}
\item \emph{Derivative in $\ee$.} By \eqref{eq:lambda}--\eqref{eq:Delta0}, $\Delta(\ee,0)=F\big(\lambda^2(\ee)\big)$ for all $\ee\in(0,1)$, where
\[
F(x):=-2\cosh\big(2\pi\sqrt x\big)=-2\sum_{n\ge0}\frac{(2\pi)^{2n}}{(2n)!}\,x^{n},\qquad \lambda^2(\ee)=\frac{9\ee^2-1}{4\D}.
\]
Since $F$ is an entire function of $x$, this representation holds on both sides of $\ee=1/3$, where $\lambda^2$ changes sign, and no branch point occurs. We have $F'(0)=-4\pi^2$, and since the numerator of $\lambda^2$ vanishes at $\ee=1/3$, where $\D=8/9$,
\[
\frac{{\rm d}\lambda^2}{{\rm d}\ee}\Big|_{\ee=\frac{1}{3}}=\frac{18\ee}{4\D}\Big|_{\ee=\frac{1}{3}}=\frac{27}{16}.
\]
Hence
\begin{equation}\label{eq:dDde}
\partial_\ee\Delta(1/3,0)=-4\pi^2\cdot\frac{27}{16}=-\frac{27\pi^2}{4}\neq0 .
\end{equation}

\item \emph{Derivative in $\nu^2$.} The fundamental matrix $\F$ solves $\F'=\mathbb A\F$, $\F(0)=\mathbb I$, and depends on $\nu^2$. Differentiating this problem with respect to $\nu^2$ by the product rule, $\Psi:=\partial_{\nu^2}\F$ solves
\[
\Psi'=\mathbb A\,\Psi+\big(\partial_{\nu^2}\mathbb A\big)\F,\qquad \Psi(0)=0,
\]
the initial value being zero because $\F(0)=\mathbb I$ does not depend on $\nu^2$. This is the equation for $\F$ with the  forcing $\boldsymbol{\mathfrak{ F}}:=(\partial_{\nu^2}\mathbb A)\F$, and it is solved by the variation-of-constants formula: the solution of $\mathbf Y'=\mathbb A\mathbf Y+\boldsymbol{\mathfrak{ F}}(\theta)$ with $\mathbf Y(0)=0$ is
\[
\mathbf Y(\theta)=\F(\theta)\int_0^{\theta}\F(s)^{-1}\boldsymbol{\mathfrak{ F}}(s)\,{\rm d}s,
\]
as one checks by direct computations. Evaluating at $\theta=2\pi$ and $\nu=0$, and writing $\F_0:=\F|_{\nu=0}$, we obtain
\[
\partial_{\nu^2}\mathbb M\big|_{\nu=0}=\F_0(2\pi)\int_0^{2\pi}\F_0(s)^{-1}\,\partial_{\nu^2}\mathbb A(s)\,\F_0(s)\,{\rm d}s .
\]
Since $\Delta=\tr\mathbb M-1$ and the trace is linear, $\partial_{\nu^2}\Delta=\tr(\partial_{\nu^2}\mathbb M)$, and the trace may be taken inside the integral. By \eqref{eq:Asys}, only the entries $(3,1)$ and $(3,2)$ of $\mathbb A$ depend on $\nu^2$, and linearly; hence
\[
\partial_{\nu^2}\mathbb A(s)=-\frac{1}{\D}\begin{pmatrix}0&0&0\\0&0&0\\ q_1(s)&q_2(s)&0\end{pmatrix}
=-\frac{1}{\D}\,\mathbf e_3\,\widetilde{\mathbf q}(s)^T,
\]
where $\mathbf e_3=(0,0,1)^T$ and $\widetilde{\mathbf q}:=(q_1,q_2,0)^T$: a matrix of rank one. For a column vector $\mathbf a$ and a row vector $\mathbf b^T$, the cyclicity of the trace gives $\tr(\mathbf a\,\mathbf b^T)=\mathbf b^T\mathbf a$. Applying this with $\mathbf a=\F_0(2\pi)\F_0(s)^{-1}\mathbf e_3$ and $\mathbf b^T=\widetilde{\mathbf q}(s)^T\F_0(s)$ yields
\begin{equation}\label{eq:dDdnu0}
\partial_{\nu^2}\Delta\big|_{\nu=0}=-\frac{1}{\D}\int_0^{2\pi}\widetilde{\mathbf q}(s)^T\,\F_0(s)\,\F_0(2\pi)\,\F_0(s)^{-1}\,\mathbf e_3\,{\rm d}s .
\end{equation}

Let us describe now the structure of $\F_0$. At $\nu=0$ the last equation in \eqref{eq:compact} reads $\widetilde{\B}'=0$, so $\widetilde{\B}$ is constant and enters the first equation, $\mathbf V'=\Msf\mathbf V+\widetilde{\B}\,\mathbf q$, as a constant forcing coefficient. Denoting by $X(\theta)$ the fundamental matrix of $\mathbf V'=\Msf\mathbf V$, the variation-of-constants formula gives
\[
\mathbf V(\theta)=X(\theta)\Big[\mathbf V(0)+\widetilde{\B}(0)\,\mathbf j(\theta)\Big],\qquad
\mathbf j(\theta):=\int_0^{\theta}X(s)^{-1}\mathbf q(s)\,{\rm d}s,
\]
that is, in the variables $(\mathbf V,\widetilde{\B})$,
\[
\F_0(\theta)=\begin{pmatrix}X(\theta)&X(\theta)\,\mathbf j(\theta)\\ 0&1\end{pmatrix},\qquad
\F_0(\theta)^{-1}=\begin{pmatrix}X(\theta)^{-1}&-\mathbf j(\theta)\\ 0&1\end{pmatrix},
\]
the inverse being checked by multiplication, since the upper right block of the product is $-X\mathbf j+X\mathbf j=0$. Hence $\F_0(s)^{-1}\mathbf e_3=(-\mathbf j(s),1)^T$, and successively
\[
\begin{aligned}
\F_0(2\pi)\begin{pmatrix}-\mathbf j(s)\\ 1\end{pmatrix}&=\begin{pmatrix}X(2\pi)\big(\mathbf j(2\pi)-\mathbf j(s)\big)\\ 1\end{pmatrix},\\[4pt]
\F_0(s)\begin{pmatrix}X(2\pi)\big(\mathbf j(2\pi)-\mathbf j(s)\big)\\ 1\end{pmatrix}
&=\begin{pmatrix}X(s)\big[X(2\pi)\big(\mathbf j(2\pi)-\mathbf j(s)\big)+\mathbf j(s)\big]\\ 1\end{pmatrix}.
\end{aligned}
\]
Since the third component of $\widetilde{\mathbf q}$ vanishes, only the first two components contribute in \eqref{eq:dDdnu0}, and
\begin{equation}\label{eq:dDdnu}
\partial_{\nu^2}\Delta\big|_{\nu=0}=-\frac1\D\int_0^{2\pi}\mathbf q(s)\cdot X(s)\Big[X(2\pi)\big(\mathbf j(2\pi)-\mathbf j(s)\big)+\mathbf j(s)\Big]\,{\rm d}s .
\end{equation}

By the construction leading to \eqref{eq:W}--\eqref{eq:Delta0}, the change of variables $\mathbf V=R_{\alpha(\theta)}\mathbf W$, with $\alpha(\theta)=\theta/2+\pi/4$, transforms $\mathbf V'=\Msf\mathbf V$ into $\mathbf W'=\mathsf K\mathbf W$, so that $\mathbf W(\theta)=E(\theta)\mathbf W(0)$ with $E(\theta):=e^{\theta\mathsf K}$. Since $\alpha(0)=\pi/4$ and the inverse of a rotation is its transpose, $\mathbf W(0)=R_{\pi/4}^T\mathbf V(0)$, and therefore
\[
X(\theta)=R_{\alpha(\theta)}E(\theta)R_{\pi/4}^T,\qquad X(\theta)^{-1}=R_{\pi/4}E(\theta)^{-1}R_{\alpha(\theta)}^T .
\]
At $\theta=2\pi$ one has $\alpha=\pi+\pi/4$, and a further rotation by $\pi$ changes the sign, $R_{\pi+\beta}=-R_\beta$; hence $X(2\pi)=-R_{\pi/4}E(2\pi)R_{\pi/4}^T$. Set
\begin{equation}\label{J}
\mathbf r(s):=R_{\alpha(s)}^T\mathbf q(s),\qquad \boldsymbol{\mathfrak{J}}(s):=\int_0^{s}E(\sigma)^{-1}\mathbf r(\sigma)\,{\rm d}\sigma,
\end{equation}
\begin{equation}\label{ys}
\mathbf y(s):=\boldsymbol{\mathfrak{J}}(s)-E(2\pi)\big(\boldsymbol{\mathfrak{J}}(2\pi)-\boldsymbol{\mathfrak{J}}(s)\big).
\end{equation}
From the expression of $X^{-1}$ we have $\mathbf j=R_{\pi/4}\boldsymbol{\mathfrak{J}}$, so that the bracket in \eqref{eq:dDdnu} equals $R_{\pi/4}\mathbf y(s)$, the minus sign in front of $E(2\pi)$ coming from $X(2\pi)$. Then $X(s)R_{\pi/4}\mathbf y=R_{\alpha(s)}E(s)\mathbf y$, and since $\mathbf q\cdot(R\mathbf v)=(R^T\mathbf q)\cdot\mathbf v$ for every rotation $R$, \eqref{eq:dDdnu} becomes
\begin{equation}\label{eq:dDdnurot}
\partial_{\nu^2}\Delta\big|_{\nu=0}=-\frac1\D\int_0^{2\pi}\mathbf r(s)\cdot E(s)\,\mathbf y(s)\,{\rm d}s .
\end{equation}

It is now convenient to introduce the following orthonormal vectors $\boldsymbol{\upsilon}_\pm:=(1,\pm1)^T/\sqrt2$; if vectors are written in coordinates, $\mathbf a=a_+\boldsymbol{\upsilon}_++a_-\boldsymbol{\upsilon}_-$, the scalar product is $\mathbf a\cdot\mathbf b=a_+b_++a_-b_-$. With $\mathsf K=\left(\begin{smallmatrix}\eta&-\varpi\\ \varpi&-\eta\end{smallmatrix}\right)$ from \eqref{eq:W}, a direct multiplication gives
\[
\mathsf K\boldsymbol{\upsilon}_+=(\eta-\varpi)\,\boldsymbol{\upsilon}_-,\qquad \mathsf K\boldsymbol{\upsilon}_-=(\eta+\varpi)\,\boldsymbol{\upsilon}_+ .
\]
To express $\mathbf r$, put $u=s/2$, so that $\alpha=u+\pi/4$, and $(r_1,r_2)=R_\alpha^T\mathbf q=(q_1\cos\alpha+q_2\sin\alpha,\ -q_1\sin\alpha+q_2\cos\alpha)$ with $\mathbf q=(\ee\sin s,\,1-\ee\cos s)$. Using $\cos\alpha-\sin\alpha=-\sqrt2\sin u$ and $\cos\alpha+\sin\alpha=\sqrt2\cos u$, together with $\cos(s-u)=\cos u$ and $\sin(s-u)=\sin u$ (because $s-u=u$), we find
\[
\begin{aligned}
r_1+r_2&=\sqrt2\,\big(q_2\cos u-q_1\sin u\big)=\sqrt2\,\big(\cos u-\ee\cos(s-u)\big)=\sqrt2\,(1-\ee)\cos u,\\
r_1-r_2&=\sqrt2\,\big(q_1\cos u+q_2\sin u\big)=\sqrt2\,\big(\sin u+\ee\sin(s-u)\big)=\sqrt2\,(1+\ee)\sin u,
\end{aligned}
\]
that is, since $r_\pm=(r_1\pm r_2)/\sqrt2$,
\begin{equation}
    \label{r}
\mathbf r(s)=\underbrace{(1-\ee)\cos\frac{s}{2}}_{:=r_+}\,\boldsymbol{\upsilon}_++\underbrace{(1+\ee)\sin\frac{s}{2}}_{:=r_-}\,\boldsymbol{\upsilon}_- .
\end{equation}

Now set $\ee=1/3$, giving $\D=8/9$, $\eta=\ee/\D=3/8$ and $\varpi=(3\ee^2-1)/(2\D)=-3/8$ by \eqref{eq:gsome}, so that
\[
\mathsf K\boldsymbol{\upsilon}_+=\frac{3}{4}\,\boldsymbol{\upsilon}_-,\qquad \mathsf K\boldsymbol{\upsilon}_-=0 .
\]
Applying $\mathsf K$ twice to each basis vector gives zero, i.e.\ $\mathsf K^2=0$, and the exponential series terminates: $E(s)=\mathbb I+s\mathsf K$. In coordinates,
\begin{equation}
    \begin{aligned}
        &E(s)(x_+,x_-)=\big(x_+,\ x_-+\frac{3}{4} s\,x_+\big),\\
&E(s)^{-1}(x_+,x_-)=\big(x_+,\ x_--\frac{3}{4} s\,x_+\big),
    \end{aligned}
\end{equation}
and in particular $E(2\pi)(x_+,x_-)=\big(x_+,\ x_-+\frac{3\pi}{2}x_+\big)$. \\
Using \eqref{r}, it follows that
\[
\mathbf r(s)=\Big(\frac{2}{3}\cos\frac{s}{2},\ \frac{4}{3}\sin\frac{s}{2}\Big),\qquad
E(s)^{-1}\mathbf r(s)=\Big(\frac{2}{3}\cos\frac{s}{2},\ \frac{4}{3}\sin\frac{s}{2}-\frac{s}{2}\,\cos\frac{s}{2}\Big).
\]

Write $\boldsymbol{\mathfrak{J}}=(\mathfrak{j}_+, \mathfrak{j}_-)$. Computing the first and second components gives:
\[
\begin{aligned}
\mathfrak{j}_+(s)&=\int_0^s\frac{2}{3}\cos\frac{\sigma}{2}\,{\rm d}\sigma=\frac{4}{3}\sin\frac{s}{2},\\
\mathfrak{j}_-(s)
&=\int_0^s\Big( \frac{4}{3}\sin\frac{\sigma}{2}-\frac{1}{2}\,\sigma\cos\frac{\sigma}{2}\Big){\rm d}\sigma=\frac{14}{3}\Big(1-\cos\frac{s}{2}\Big)-s\sin\frac{s}{2} .
\end{aligned}
\]
In particular $\boldsymbol{\mathfrak{J}}(2\pi)=\big(0,\frac{28}{3}\big)$, and
\[
\boldsymbol{\mathfrak{J}}(2\pi)-\boldsymbol{\mathfrak{J}}(s)=\Big(-\frac{4}{3}\sin\frac{s}{2},\ \ \frac{14}{3}+\frac{14}{3}\cos\frac{s}{2}+s\sin\frac{s}{2}\Big).
\]
Applying $E(2\pi)$, which adds $\frac{3\pi}{2}$ times the first component to the second, and subtracting the result from $\boldsymbol{\mathfrak{J}}(s)$ according to \eqref{ys}, we obtain
\[
\mathbf y(s)=\Big(\frac{8}{3}\sin\frac{s}{2},\ \ -\frac{28}{3}\cos\frac{s}{2}-2s\sin\frac{s}{2}+2\pi\sin\frac{s}{2}\Big),
\]
and applying $E(s)$, which adds $\frac{3}{4} s\,y_+=2s\sin\frac{s}{2}$ to the second component,
\[
E(s)\,\mathbf y(s)=\Big(\frac{8}{3}\sin\frac{s}{2},\ \ 2\pi\sin\frac{s}{2}-\frac{28}{3}\cos\frac{s}{2}\Big).
\]
 Hence, using the duplication formulas $\sin\frac{s}{2}\cos\frac{s}{2}=\frac{1}{2}\sin s$, $\sin^2\frac{s}{2}=\frac{1}{2}(1-\cos s)$, we have
\[
\mathbf r(s)\cdot E(s)\,\mathbf y(s)=-\frac{16}{3}\sin s+\frac{4\pi}{3}\,(1-\cos s).
\]
The integrals of $\sin s$ and $\cos s$ over $[0,2\pi]$ vanish, so the integral in \eqref{eq:dDdnurot} equals $\frac{4\pi}{3}\cdot2\pi=\frac{8\pi^2}{3}$, and \eqref{eq:dDdnurot} with $\D=8/9$ gives
\begin{equation}\label{eq:dDdnu2}
\partial_{\nu^2}\Delta(1/3,0)=-\frac98\cdot\frac{8\pi^2}{3}=-3\pi^2 .
\end{equation}
In conclusion, since $\Delta(1/3,0)+2=0$ and $\partial_\ee\Delta(1/3,0)\neq0$, the implicit function theorem applies, and
\begin{equation}
\ee_c'(0)=-\frac{-3\pi^2}{-\frac{27\pi^2}{4}}=-\frac49.
\end{equation}
As $\ee_c$ is real-analytic in $\nu^2$, its Taylor expansion reads $\ee_c(\nu^2)=\frac13-\frac49\nu^2+\mathcal{O}\big((\nu^2)^2\big)$, which is \eqref{eq:ec}.\\
Finally, near $(1/3,0)$ we have $\Delta$ close to $-2$, hence $\Delta<2$, and $\partial_\ee\Delta<0$ by continuity of the derivative, so that $\Delta$ is strictly decreasing in $\ee$ at fixed $\nu^2$. Consequently $\Delta>-2$, i.e.\ $|\Delta|<2$, for $\ee<\ee_c(\nu^2)$, and $\Delta<-2$, i.e.\ $|\Delta|>2$, for $\ee>\ee_c(\nu^2)$: in that neighbourhood the level is unstable if and only if $\ee>\ee_c(\nu^2)$.
\end{itemize}
\item[(iii)]
Since $\mathbb A$ is regular at $\ee=0$, $\Delta$ is analytic, and in particular continuous, up to $\ee=0$. At $\ee=0$ one has $\Msf\equiv0$ and $\mathbf q\equiv(0,1)^T$, so that \eqref{eq:Asys} has the constant matrix
\[
\mathbb A_0=\begin{pmatrix}0&0&0\\0&0&1\\0&-\nu^2&0\end{pmatrix},
\qquad\text{i.e.}\qquad
\mathfrak{U}_1'=0,\quad \mathfrak{U}_3''=-\nu^2\,\mathfrak{U}_3 .
\]
Physically, $\mathfrak{U}_3$ performs a buoyancy oscillation of frequency $\nu$ in $\theta$, that is of frequency $\nu\Omega=\mathcal N$ in time, while $\mathfrak{U}_1$ is frozen. Hence $\mathbb M=e^{2\pi\mathbb A_0}$ is block diagonal, with the entry $1$ for $\mathfrak{U}_1$ and the block
\[
\begin{cases}
\begin{pmatrix}
\cos (2\pi\nu) & \nu^{-1}\sin (2\pi\nu)\\
-\nu\sin (2\pi\nu) & \cos (2\pi\nu)
\end{pmatrix},
& \quad\text{if}\quad\nu^2>0,  \\[1.2em]
\begin{pmatrix}
\cosh (2\pi|\nu|) & |\nu|^{-1}\sinh (2\pi|\nu|)\\
|\nu|\sinh (2\pi|\nu|) & \cosh (2\pi|\nu|)
\end{pmatrix},
& \quad\text{if}\quad \nu^2<0.
\end{cases}
\]
 Therefore $\tr\mathbb M=1+2\cos(2\pi\nu)$ or $\tr\mathbb M=1+2\cosh(2\pi|\nu|)$ , and
\[
\Delta(0,\nu^2)=2\cos(2\pi\nu)\quad(\nu^2\ge0),\qquad \Delta(0,\nu^2)=2\cosh(2\pi|\nu|)\quad(\nu^2<0),
\]
the two expressions agreeing, with value $2$, at $\nu=0$. Two consequences follow by continuity of $\Delta$.

If $\nu^2<0$, then $\Delta(0,\nu^2)>2$, hence $\Delta(\ee,\nu^2)>2$ for all sufficiently small $\ee\geq0$: a statically unstable layer is unstable for arbitrarily small steepness, with growth rate $\sigma\to\frac{\Omega}{2\pi}\arcosh\big(\cosh2\pi|\nu|\big)=\Omega|\nu|$ as $\ee\to0$, the classical convective growth rate (see e.g. \cite{Chandrasekhar,Vallis}).

If $\nu^2>0$ and $2\nu\notin\mathbb Z$, then $|\Delta(0,\nu^2)|=|2\cos2\pi\nu|<2$ strictly, hence $|\Delta|<2$ on a whole neighbourhood of $(0,\nu^2)$; this neighbourhood does not meet $\mathcal S$, and $(0,\nu^2)$ does not belong to the closure of $\mathcal S$. The only points of the axis $\ee=0$ with $\nu^2>0$ that may belong to that closure are therefore those with $|2\cos2\pi\nu|=2$, i.e.\ $\nu=n/2$ with $n\in\mathbb N$.
\end{enumerate}
\end{proof}

\paragraph{Numerics}
We now turn to the numerical results. Figure \ref{fig:stability} shows the instability diagram for the problem at hand. The unstable region $\mathcal S$ in the $(\ee,\nu^2)$-plane is coloured according to the growth rate \eqref{eq:sigma}, computed from the Hill discriminant of \eqref{eq:Hill}. The instability region found for $\nu=0$ persists for $\nu^2\neq0$; stable stratification lowers the critical steepness, consistently with the asymptotic prediction \eqref{eq:ec}, and connects the main instability region with the subharmonic resonance tongue issuing from $\nu=1/2$. For steep waves ($\ee\gtrsim0.35$), it slightly reduces the growth rate (see Fig. \ref{fig:rate}).\\
\noindent
The discriminant was computed by integrating the Hill system \eqref{eq:Hillsys} over one period using the classical fourth-order Runge--Kutta scheme with $N_\theta=10^4$ steps, simultaneously on a $1000\times1000$ grid in the $(\ee,\nu^2)$-plane. As an independent check, $\Delta$ was also obtained from the full $3\times3$ system \eqref{eq:Asys} using an adaptive eighth-order Dormand--Prince integrator with relative tolerance $10^{-12}$. The two computations agree to within the integration tolerance.\\
\noindent
A more quantitative comparison with the asymptotic result \eqref{eq:ec} is provided in Table~\ref{tab}, which reports the neutral steepness $\ee_c(\nu^2)$ obtained by solving $\Delta(\ee,\nu^2)=-2$ numerically, together with the corresponding asymptotic prediction. The numerical roots were obtained by direct integration of either the full system \eqref{eq:Asys} or the equivalent Hill equation \eqref{eq:Hill}.
\begin{table}[h]
\centering
\caption{Neutral steepness $\ee_c(\nu^2)$: numerically computed root of $\Delta(\ee,\nu^2)=-2$ versus the asymptotic formula \eqref{eq:ec}.}
\label{tab}
\begin{tabular}{@{}cccc@{}}
\toprule
$\nu^2$ & $\ee_c$ (numerical) & $\frac{1}{3}-\frac{4}{9}\nu^2$ & relative error\\
\midrule
$0.001$ & $0.33289$ & $0.33289$ & $0.0003\%$\\
$0.01$ & $0.32878$ & $0.32889$ & $0.03\%$\\
$0.03$ & $0.31900$ & $0.32000$ & $0.3\%$\\
$0.05$ & $0.30816$ & $0.31111$ & $1\%$\\
$0.1$ & $0.27478$ & $0.28889$ & $5\%$\\
$1/6$ & $0.20560$ & $0.25926$ & $26\%$\\
$1/5$ & $0.14898$ & $0.24444$ & $64\%$\\
$0.24$ & $0.03804$ & $0.22667$ & ---\\
$0.248$ & $0.00793$ & $0.22311$ & ---\\
\bottomrule
\end{tabular}
\end{table}
The agreement is excellent for $\nu^2\lesssim10^{-2}$, but progressively deteriorates as $\nu^2$ approaches $1/4$, where the neutral curve terminates at the tip $(\ee,\nu^2)=(0,1/4)$ of the subharmonic tongue (see Fig. \ref{fig:stability}). Numerically, the approximation \eqref{eq:ec} can be improved to
\begin{equation}\label{eq:ec+}
\ee_c(\nu^2)\approx\frac{1}{3}-\frac{4}{9}\nu^2-1.02\nu^4,
\end{equation}
which reduces the relative error from $1\%$ to $0.13\%$ at $\nu^2=0.05$, and from $5\%$ to $1.4\%$ at $\nu^2=0.1$.\\
\\
Observing Fig. \ref{fig:stability}, one can note that, for $\nu^2<0$, namely unstably stratified layer, there is a white band that separates two unstable regions. Indeed, this band separates two instabilities of different nature, which cannot be continuously connected without passing through stability. For small steepness the layer is convectively unstable: $\Delta>2$, and the multipliers $\mu^{\pm1}$ are real and positive. For steep waves the instability is the wave-induced subharmonic resonance: $\Delta<-2$, and $\mu^{\pm1}$ are real and negative. Since $\mu\mu^{-1}=1$, a real pair cannot change sign by passing through zero; it has to leave the real axis, and non-real multipliers lie on the unit circle (Corollary~\ref{cor:discriminant}). As $\ee$ increases across the band, the pair $\mu^{\pm1}=e^{\pm i\varphi}$ therefore travels along the unit circle, from the double value $+1$ at $\ee=\ee_a$ ($\varphi=0$) to the double value $-1$ at $\ee=\ee_b$ ($\varphi=\pi$).\\
This observation can be made precise. Let $\nu^2<0$ and suppose that $\Delta(\ee_1,\nu^2)<-2$ for some $\ee_1\in(0,1)$. Since $\Delta(\ee,\nu^2)$ is continuous in $\ee$ and, by Theorem~\ref{prop:strat}(iii),
$$
\Delta(0,\nu^2)=2\cosh(2\pi|\nu|)>2,
$$
there must exist two values $0<\ee_a<\ee_b<\ee_1$ such that
$$
\Delta(\ee_a,\nu^2)=2,\qquad
\Delta(\ee_b,\nu^2)=-2,
$$
with
$$
|\Delta(\ee,\nu^2)|<2,\qquad \ee_a<\ee<\ee_b.
$$
Thus, the interval $(\ee_a,\ee_b)$ is a stable band separating the two instability regions. In particular, the two band edges correspond to the transition $\mu^{\pm1}=+1$ at $\ee=\ee_a$ and $\mu^{\pm1}=-1$ at $\ee=\ee_b$.\\
The existence of such a band is guaranteed for sufficiently small $|\nu^2|$. Indeed, fixing any $\ee_1>1/3$, one has $\Delta(\ee_1,0)=-2\cosh(2\pi\lambda)<-2$ by \eqref{eq:Delta0}, and continuity with respect to $\nu^2$ implies that $\Delta(\ee_1,\nu^2)<-2$ for all sufficiently small $|\nu^2|$. The upper edge $\ee_b$ is then given by the neutral curve $\Delta=-2$ already obtained in Theorem~\ref{prop:strat}(ii), and therefore
$$
\ee_b=\frac13-\frac49\nu^2+\mathcal{O}(\nu^4)
=\frac13+\frac49|\nu^2|+\mathcal{O}(\nu^4).
$$
The scaling of the lower edge can be obtained from the behaviour of $\Delta$ near $(\ee,\nu^2)=(0,0)$. Since $\Delta$ is even in $\ee$, it is analytic in $x=\ee^2$ and $y=\nu^2$ near the origin. From \eqref{eq:lambda}--\eqref{eq:Delta0}, for $y=0$,
$$
\Delta=2-16\pi^2x^2+\mathcal{O}(x^3),
$$
while Theorem~\ref{prop:strat}(iii) gives, along $x=0$,
$$
\Delta=2-4\pi^2y+\mathcal{O}(y^2).
$$
Hence, near the origin,
$$
\Delta-2=-4\pi^2y-16\pi^2x^2
+\mathcal{O}(xy)+\mathcal{O}(y^2)+\mathcal{O}(x^3).
$$
Since $\partial_y(\Delta-2)|_{(0,0)}=-4\pi^2\neq0$, the implicit function theorem gives a unique local branch of the curve $\Delta=2$, of the form $y=g(x)$, with
$$
g(x)=-4x^2+\mathcal{O}(x^3).
$$
Consequently, for $\nu^2<0$,
$$
\nu^2=-4\ee_a^4\bigl(1+\mathcal{O}(\ee_a^2)\bigr),
$$
or equivalently
$$
\ee_a=\Bigl(\frac{|\nu^2|}{4}\Bigr)^{1/4}
\bigl(1+\mathcal{O}(|\nu|)\bigr).
$$
Thus, for weakly unstable stratification, the stable band has edges
$$
\ee_a=\Bigl(\frac{|\nu^2|}{4}\Bigr)^{1/4}
\bigl(1+\mathcal{O}(|\nu|)\bigr),
\qquad
\ee_b=\frac13+\frac49|\nu^2|+\mathcal{O}(\nu^4).
$$
The lower edge has a simple meaning. By \eqref{eq:Qmean}, the mean stiffness of the Hill equation vanishes when $\nu^2=-4\ee^4/\D^2=-4\ee^4+\mathcal{O}(\ee^6)$, which is the leading-order relation defining $\ee_a$. In other words, the band begins where the mean restoring effect of the wave, $4\ee^4/\D^3$, balances the negative buoyancy stiffness $\nu^2/\D$ of the statically unstable layer: the orbital motion dynamically stabilises the convective mode, much as a rapid vertical oscillation stabilises Kapitza's inverted pendulum \cite{Kapitza1951,LeviWeckesser1995}. The upper edge is the onset of the wave-induced parametric resonance.


%

\section{Discussion}\label{discussion}
\begin{figure}[h!]
    \centering
    \includegraphics[width=0.85\linewidth]{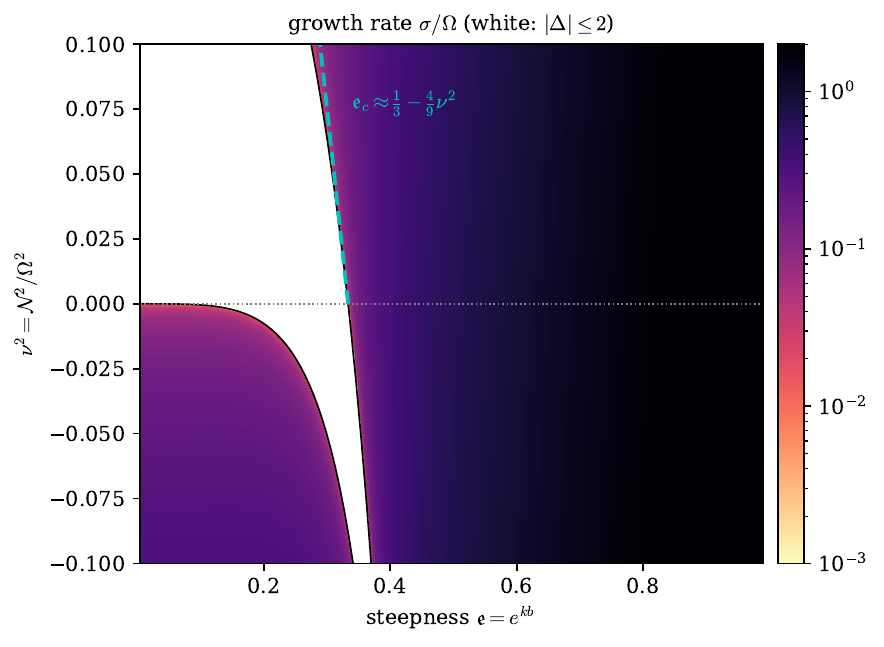}
    \caption{Instability diagram for $|\nu^2|\leq 0.1$}
    \label{fig:stab_small}
\end{figure}

We have presented an instability analysis of the nonlinear solution for mountain waves proposed in \cite{ConstantinMountain2023}. Adopting the short-wavelength instability approach of \cite{LH} and \cite{LebovitzL} for compressible flows, we have been able to transform the study of the instability of a complicated flow, given in Lagrangian coordinates, into the analysis of a system of ODEs for the amplitude of the short-wavelength perturbation. Compared with the incompressible case (see, e.g., \cite{CG,Kruse,Kruse2,mioInstability}), compressibility and stratification introduce a coupling between the velocity amplitude $\A$ and the entropy amplitude $\B$, making the analysis considerably more involved.\\
The linear system for such a compressible trochoidal flow, whose coefficients are periodic in time, cannot be transformed, in general, into a system with constant coefficients by a rotation of the amplitude space, as is the case for incompressible flows. Its analysis therefore relies on Floquet theory \cite{YakubovichStarzhinskii1975}. Volume preservation, together with the conservation of $\mathcal I$, the leading-order WKB amplitude of Ertel's potential vorticity, forces the Floquet multipliers to be of the form ${1,\mu,\mu^{-1}}$. The multiplier $1$ is associated with the conserved quantity $\mathcal I$, while the pair $\mu^{\pm1}$ governs stability. In particular, a sufficient condition for the emergence of an exponentially growing solution is $|\Delta|>2$, where $\Delta=\mu+\mu^{-1}$ is the trace of the monodromy matrix minus one. In this case, one of $\mu$ and $\mu^{-1}$ is real and different from $\pm1$, and the amplitude of the perturbation grows exponentially in time.\\
Moreover, the growing and decaying modes carry no potential vorticity, and on the corresponding invariant plane the problem reduces to a Hill equation whose Floquet discriminant equals $\Delta$. This reduction not only provides a simpler characterization of the instability, but also allows us to identify its nature. In the regime of interest, $\Delta<-2$, so that the unstable mode changes sign at every orbit and the instability is subharmonic. \\
The stability of each level of the flow depends on two dimensionless parameters only: the steepness $\ee=e^{kb}$ of the trochoidal isolines, as in the incompressible case, and $\nu^2=\mathcal N^2/\Omega^2$, the squared ratio between the Brunt--V\"ais\"al\"a frequency and the orbital frequency of the air parcels. The uniform velocities do not enter, and compressibility enters only through $\nu^2$. For neutral stratification, $\nu=0$, the problem can be solved explicitly: in the frame rotating with the principal axes of strain, which turn at half the orbital rate, the system becomes autonomous, and instability occurs if and only if $\ee>1/3$, with growth rate $\sigma=\tfrac{\Omega}{2}\sqrt{(9\ee^2-1)/(1-\ee^2)}$ (Theorem~\ref{thm:neutral}). This is the threshold found in \cite{Leblanc2004} for Gerstner's water wave, whose particle kinematics coincide with those of \eqref{solution}; physically, it expresses the fact that the strain exceeds the effective rotation felt by the perturbation in the co-rotating frame. The neutral case is realised by an explicit isentropic basic state, which is therefore unstable whenever the top of the laminar layer satisfies $b_0>-\ln3/k$. For weakly stable stratification we have shown that, near $(\ee,\nu^2)=(1/3,0)$, a level is unstable if and only if
\begin{equation}\label{criterion}
\ee>\ee_c(\nu^2)=\frac13-\frac49\,\nu^2+\mathcal{O}(\nu^4),
\end{equation}
so that stable stratification slightly lowers the critical steepness (Theorem~\ref{prop:strat}). \\
For general $(\ee,\nu^2)$ the discriminant has been computed numerically, leading to the stability diagram of Fig.~\ref{fig:stability}. Besides the main instability region, the diagram displays resonance tongues issuing from $\nu=n/2$; these, however, require buoyancy frequencies comparable to the orbital frequency and lie outside the atmospheric range. For statically unstable layers, $\nu^2<0$, the flow is unstable already for arbitrarily small steepness, with growth rate tending to the classical convective value $|\mathcal N|$ as $\ee\to0$. Fig.~\ref{fig:stability} shows an interesting feature: for $\nu^2<0$ not every steepness leads to instability, since a thin stable band separates the convective instability, for which $\Delta>2$, from the wave-induced one, for which $\Delta<-2$. The existence of this band follows from the continuity of $\Delta$, which must cross the interval $[-2,2]$ between the two regions, while its precise location and width deserve further analysis.\\
It is also interesting to provide some physical estimates for the problem under consideration. Let us recall that the stratification parameter $\nu^2$ is defined as
\begin{equation}\label{nu quadro}
    \nu^2=\frac{\mathcal{N}^2}{\Omega^2},
\end{equation}
where $\Omega^2=gk$ and $\mathcal{N}^2$ is the squared non-dimensional Brunt--V\"ais\"al\"a frequency in Lagrangian coordinates (the Eulerian one being $N^2=\mathcal N^2 b_z$). Moreover, \eqref{nu quadro}, being the ratio of two squared frequencies, has the same value also if $\Omega^2$ and $\mathcal{N}^2$ are replaced by their dimensional versions $\Omega'^2$ and $\mathcal{N}'^2$ (here $'$ denotes dimensional quantities). By \eqref{jacob}, $b_z=\frac{1-\ee\cos\theta}{1-\ee^2}$, and $\nu^2=N'^2/(b_z\,\Omega'^2)$. Since $\ee\in(0,1)$, we have
\begin{equation}\label{nu bounds}
    \frac{1}{1+\ee}\leq b_z\leq \frac{1}{1-\ee} \quad\Longrightarrow\quad 0<|\nu^2| \leq 2\,\frac{|N'^2|}{\Omega'^2}\quad \text{and}\quad \mathrm{sign}(\nu^2)=\mathrm{sign}(N'^2);
\end{equation}
in particular, since $b_z$ is of order one, $\nu^2$ is of the order of $N'^2/\Omega'^2$. In the troposphere the buoyancy frequency is typically $N'\approx10^{-2}\,{\rm s^{-1}}$, i.e.\ $N'^2\approx10^{-4}\,{\rm s^{-2}}$ \cite{20}. On the other hand, $\Omega'^2=g'k'=2\pi g'/\lambda'$, where $\lambda'=2\pi/k'$ is the horizontal wavelength, so that $N'^2/\Omega'^2=N'^2\lambda'/(2\pi g')$ grows linearly with $\lambda'$. In linear theory, mountain waves propagate vertically for sufficiently long horizontal wavelengths, which may reach several tens of kilometres, e.g.\ $\lambda'=40\,$km \cite{durran,durran2}. For such a wavelength, \eqref{nu bounds} gives $|\nu^2|\leq N'^2\lambda'/(\pi g')\approx0.13$, so that the physically relevant values of $\nu^2$ are at most of order $10^{-1}$.The physically relevant values of $\nu^2$ are therefore between $10^{-2}$ and $10^{-1}$ in order of magnitude, and Fig.~\ref{fig:stab_small} shows the analogue of Fig.~\ref{fig:stability} restricted to $|\nu^2|\leq0.1$. 
\noindent

\noindent
Short‑wavelength instabilities in inviscid incompressible flows are believed to be responsible for the breakdown of two‑dimensional coherent structures into three‑dimensional chaotic motion (see \cite{lifschitz} and references therein). In the present, compressible setting, the same mechanism is expected to persist. Our analysis therefore suggests that the coherent two‑dimensional solution \eqref{solution} may be destabilized by short-wavelength perturbations. The ensuing growth would tilt and stretch material structures, promote three‑dimensionality through vortex‑line bending, and facilitate energy transfer to smaller scales (similarly to the mechanism observed for Kelvin-Helmholtz waves in \cite{Klaassen}), ultimately seeding turbulence in the upper troposphere and near the tropopause. This scenario is consistent with the broader view that (short‑wavelength) instabilities provide a generic route from quasi‑two‑dimensional organized flow to fully three‑dimensional fluid motion (see e.g. \cite{PW82,Or, corcos1, corcos2,Rob, P86, M_etal}).

Finally, the motion considered here is dry and adiabatic, i.e.\ it involves neither latent heating nor other diabatic effects, which necessarily simplifies the thermodynamics of the problem. We aim in future works to extend such analysis to more complicated flows (e.g. those in  \cite{henry}) or to those that account for precipitations, such as \cite{Lyons2025,McCL}.\\

\vspace{.51cm}
\paragraph*{\bf Acknowledgments} The author would like to thank the reviewers for their very useful comments.
\paragraph*{\bf Declaration of interests}
The author reports no conflict of interest.

\paragraph*{\bf Data availability statement}
No data have been used or produced for this work.

\paragraph*{\bf Code availability statement}
The code used to generate Figs.~\ref{fig:stability}, \ref{fig:rate} (and \ref{fig:stab_small}) is available in the Supplementary Material and can also be obtained from the author upon request.

\paragraph*{\bf Author ORCID}
\href{https://orcid.org/0009-0008-5454-0922}{orcid.org/0009-0008-5454-0922}.

\appendix

\section{Ertel's theorem and the conserved quantity}\label{app:PV}
Proposition~\ref{prop:PV} was established by direct computation, which shows that $\mathcal I$ is conserved but does not explain its physical meaning. The underlying mechanism is Ertel's theorem, applied to the rapidly oscillating perturbations considered here. Ertel's theorem states that the potential vorticity
\begin{equation}
q_E:=\frac{\boldsymbol\omega\cdot\nabla \mathfrak{f}}{\rho},
\qquad
\boldsymbol\omega=\nabla\times\bu,
\end{equation}
is materially conserved, $Dq_E/Dt=0$, whenever $\mathfrak{f}$ is a materially advected scalar and the corresponding baroclinic contribution vanishes (see e.g.\ \cite{Muller1995, Viudez1999}). In the present setting, the natural choice is $\mathfrak{f}=\ln\Theta$, where $\Theta=TP^{-\kappa}$ is the potential temperature. The first law gives $D(\ln\Theta)/Dt=0$, while
\begin{equation}
\nabla\ln\Theta=\gamma^{-1}\nabla\ln P-\nabla\ln\rho\qquad \Longrightarrow\qquad
(\nabla\rho\times\nabla P)\cdot\nabla\ln\Theta=0.
\end{equation}
Thus Ertel's theorem applies to $\mathfrak{f}=\ln\Theta$. The basic flow has vanishing potential vorticity. Indeed, its vorticity (cf.\ equation \eqref{vorticity}) is directed along $\mathbf e_2=(0,1,0)$, whereas $\Theta=\Theta(b)$ and $b_y=0$, so that
\begin{equation}\label{eq:PVbasic}
q_E^{\rm basic}:=
\frac{\boldsymbol\omega\cdot(\nabla\ln\Theta)}{\rho}=0.
\end{equation}
We perturb $\rho$, $\boldsymbol\omega$ and $\ln\Theta$ by $\varrho$, $\delta\boldsymbol\omega=\nabla\times\uu$ and $\delta\ln\Theta$, so that the potential vorticity $q_E$ is given by
\begin{equation}
    \begin{aligned}q_E&=\frac{(\boldsymbol\omega+\delta\boldsymbol\omega)\cdot(\nabla\ln\Theta+\nabla\delta\ln\Theta)}{(\rho+\varrho)}\\
    &=q_E^{\rm basic}+\frac1\rho\left(
\delta\boldsymbol\omega\cdot\nabla\ln\Theta
+
\boldsymbol\omega\cdot\nabla\delta\ln\Theta
\right)-\frac{\varrho}{\rho}q_E^{\rm basic}+ \text{higher order terms}.\end{aligned}
\end{equation}
Using $q_E^{\rm basic}=0$ we obtain
\begin{equation}
\delta q_E
=
\frac1\rho\left(
\delta\boldsymbol\omega\cdot\nabla\ln\Theta
+
\boldsymbol\omega\cdot\nabla\delta\ln\Theta
\right),
\end{equation}
and linearising $Dq_E/Dt=0$ gives $D\,\delta q_E/Dt=-\uu\cdot\nabla q_E^{\rm basic}=0$, where $D/Dt$ is the material derivative of the basic flow.

For the short-wavelength perturbations \eqref{up}, with phase gradient $\nabla\Phi=\boldsymbol\xi$, the leading contribution to $\delta q_E$ is of order $\e^{-1}$:
\begin{equation}\label{eq:dqE}
\delta q_E
=
\frac{i}{\e\rho}
\left[
(\boldsymbol\xi\times\A)\cdot\nabla\ln\Theta
+
(\boldsymbol\omega\cdot\boldsymbol\xi)\,
\widehat{\delta\ln\Theta}
\right]e^{i\Phi/\e}
+\mathcal{O}(1),
\end{equation}
where $\widehat{\delta\ln\Theta}$ denotes the leading-order amplitude of $\delta\ln\Theta$.

Recall that $\boldsymbol\xi=(0,1,0)$ and $\mathfrak{U}_2=0$, giving $\boldsymbol\xi\times\A=(\mathfrak{U}_3,0,-\mathfrak{U}_1)$: the perturbation vorticity lies in the $(x,z)$ plane and is the velocity amplitude rotated by $-\pi/2$, that is $-J\mathbf V$. Since $\Theta=\Theta(b)$ with ${\rm d}\ln\Theta/{\rm d}b=k\nu^{2}$ by \eqref{eq:nuphys}, and $\nabla b=\mathbf q/\D$,
\begin{equation}
\nabla\ln\Theta=\frac{k\nu^{2}}{\D}\,(q_1,0,q_2),
\end{equation}
so that, using $J^{T}=-J$ and $J\mathbf q=J^{2}\mathbf p=-\mathbf p$,
\begin{equation}\label{eq:term1}
(\boldsymbol\xi\times\A)\cdot\nabla\ln\Theta
=-\frac{k\nu^{2}}{\D}\,(J\mathbf V)\cdot\mathbf q
=\frac{k\nu^{2}}{\D}\,\mathbf V\cdot J\mathbf q
=-\frac{k\nu^{2}}{\D}\,\mathbf p\cdot\mathbf V ,
\end{equation}
or explicitly $\mathfrak{U}_3q_1-\mathfrak{U}_1q_2=\ee\sin\theta\,\mathfrak{U}_3-(1-\ee\cos\theta)\mathfrak{U}_1=-\mathbf p\cdot\mathbf V$. It is worth noting that this contribution is proportional to $\nu^{2}$, and hence exists only in a stratified layer.\\
\noindent
We now analyse the second term in \eqref{eq:dqE}. From $\ln\Theta=\gamma^{-1}\ln P-\ln\rho$ we get $\delta\ln\Theta=\fp/(\gamma P)-\varrho/\rho$, and the Eckart variables \eqref{eckart variables} together with $\gamma P=\rho\cs^{2}$ give $\fp/(\gamma P)=\nn/\cs$ and $\varrho/\rho=(\m+\nn)/\cs$. The variable $\nn$ cancels exactly---which is precisely what the Eckart transform is designed to do---and, by \eqref{eq:rescale},
\begin{equation}
\delta\ln\Theta=-\frac{\m}{\cs},\qquad
\widehat{\delta\ln\Theta}=-\frac{\B}{\cs}=-\frac{\widetilde{\B}}{c}.
\end{equation}
An immediate computation, using $\omega_2$ from \eqref{eq:antisym}, gives
\begin{equation}\label{eq:term2}
(\boldsymbol\omega\cdot\boldsymbol\xi)\,\widehat{\delta\ln\Theta}
=\Big(-\frac{2kc\,\ee^{2}}{\D}\Big)\Big(-\frac{\widetilde{\B}}{c}\Big)
=\frac{2k\ee^{2}}{\D}\,\widetilde{\B}.
\end{equation}
This contribution is proportional to the basic vorticity, and hence exists only because the wave is rotational.\\
\noindent
Adding \eqref{eq:term1} and \eqref{eq:term2} in \eqref{eq:dqE},
\begin{equation}
\delta q_E=\frac{i}{\e\rho}\frac{k}{\D}\Big(-\nu^{2}\,\mathbf p\cdot\mathbf V+2\ee^{2}\widetilde{\B}\Big)e^{i\Phi/\e}+\mathcal{O}(1)
=-\frac{ik}{\e\rho\D}\,\mathcal I\,e^{i\Phi/\e}+\mathcal{O}(1).
\end{equation}
Since $D(e^{i\Phi/\e})/Dt=\frac{i}{\e}({D\Phi}/{Dt})e^{i\Phi/\e}=0$, the $\mathcal{O}(\e^{-1})$ part of $D\,\delta q_E/Dt=0$ yields $D\big(\mathcal I/(\rho\D)\big)/Dt=0$; and $\rho$ and $\D$ are functions of $b$ alone, hence constant along particle paths. Therefore $\mathcal I$ is constant, which is exactly the statement of Proposition~\ref{prop:PV}.\\
Two limiting cases confirm the interpretation: for $\nu=0$ the layer is isentropic, the first contribution disappears and $\mathcal I=-2\ee^{2}\widetilde{\B}$ reduces to the conservation of the entropy perturbation; for $\ee\to0$ the wave carries no vorticity, the second contribution disappears and $\mathcal I\to\nu^{2}\mathfrak{U}_1$, which is conserved because $\mathfrak{U}_1'=0$ in that limit.

\end{document}